\documentclass[12pt,preprint]{emulateapj}

\newcommand{\beqarr}{\begin{eqnarray}}
\newcommand{\eeqarr}{\end{eqnarray}}
\newcommand{\beq}{\begin{equation}}
\newcommand{\eeq}{\end{equation}}

\def\simlt{\mathrel{\spose{\lower 3pt\hbox{$\mathchar"218$}}
     \raise 2.0pt\hbox{$\mathchar"13C$}}}
\def\simgt{\mathrel{\spose{\lower 3pt\hbox{$\mathchar"218$}}
     \raise 2.0pt\hbox{$\mathchar"13E$}}}

\usepackage[export]{adjustbox}

\shorttitle{Quasar Variability in the Mid-Infrared}
\shortauthors{Koz{\l}owski et al.}

\begin{document}


\title{Quasar Variability in the Mid-Infrared}

\author
{
Szymon~Koz{\l}owski\altaffilmark{1},
Christopher~S.~Kochanek\altaffilmark{2,3},
Matthew~L.~N.~Ashby\altaffilmark{4},
Roberto~J.~Assef\altaffilmark{5},\\
Mark~Brodwin\altaffilmark{6},
Peter~R.~Eisenhardt\altaffilmark{7},
Buell~T.~Jannuzi\altaffilmark{8},
Daniel~Stern\altaffilmark{7}\\
}

\altaffiltext{1}{Warsaw University Observatory, Al. Ujazdowskie 4, 00-478 Warsaw, Poland; e-mail:  
{\tt simkoz@astrouw.edu.pl}}
\altaffiltext{2}{Department of Astronomy, The Ohio State University, 140 West 18th Avenue, Columbus, OH 43210, USA; e-mail:  
{\tt kochanek.1@osu.edu}}
\altaffiltext{3}{The Center for Cosmology and Astroparticle Physics, The Ohio State University, Columbus, OH 43210, USA}
\altaffiltext{4}{Harvard-Smithsonian Center for Astrophysics, 60 Garden Street, Cambridge, MA 02138, USA}
\altaffiltext{5}{N{\'u}cleo de Astronom{\'i}a de la Facultad de Ingenier{\'i}a, Universidad Diego Portales, Av. Ej{\'e}rcito Libertador 441, Santiago, Chile}
\altaffiltext{6}{Department of Physics and Astronomy, University of Missouri, Kansas City, MO 64110, USA}
\altaffiltext{7}{Jet Propulsion Laboratory, California Institute of Technology, 4800 Oak Drive, Pasadena, CA 91109, USA}
\altaffiltext{8}{Steward Observatory, University of Arizona, Tucson, AZ 85721, USA}


\begin{abstract}
The Decadal IRAC Bo{\"o}tes Survey (DIBS) is a mid-IR variability survey of the $\sim$9 sq. deg. of the NDWFS Bo{\"o}tes Field and
extends the time baseline of its predecessor, the {\it Spitzer} Deep, Wide-Field Survey (SDWFS), from 4 to 10 years.
The {\it Spitzer Space Telescope} visited the field five times between 2004 and 2014 at 3.6 and 4.5 microns.
We provide the difference image analysis photometry for a half a million mostly extragalactic sources.
In mid-IR color-color plane, sources with quasar colors constitute the largest variability class (75\%),
16\% of the variable objects have stellar colors and the remaining 9\% have the colors of galaxies. 
Adding the fifth epoch doubles the number of variable AGNs for the same false positive rates as in SDWFS, or increases 
the number of sources by 20\% while decreasing the false positive rates by factors of 2-3 for the same variability amplitude.
We quantify the ensemble mid-IR variability of $\sim$1500 spectroscopically confirmed AGNs using single power-law structure functions, 
which we find to be steeper (index $\gamma\approx0.45$) than in the optical ($\gamma\approx0.3$),
leading to much lower amplitudes at short time-lags. This provides evidence for large emission regions, 
smoothing out any fast UV/optical variations, as the origin of infrared quasar variability. 
The mid-IR AGN structure function slope $\gamma$ seems to be uncorrelated with both the luminosity and rest-frame wavelength, 
while the amplitude shows an anti-correlation with the luminosity and a correlation with the rest-frame wavelength.

\end{abstract}
\vspace{2.2cm}



\section{Introduction}
\label{sec:intro}

There is increasing evidence that understanding the evolution of active galactic nuclei\footnote{We will use words ``quasar'' and ``AGN'' interchangeably throughout this paper.} (AGNs) 
 is crucial for understanding galaxy formation and evolution (e.g., \citealt{2006ApJS..163....1H,2008ApJ...676...33D,2013ARA&A..51..511K}). 
This has driven a rapid growth in attempts to comprehensively characterize the properties and demographics of AGNs. The {\it Spitzer Space Telescope}
has played a significant role in this effort through the development of mid-IR AGN selection
methods (e.g., \citealt{2004ApJS..154..166L,2005ApJ...631..163S,2007ApJ...671.1365H,2008ApJ...687..111D,2009ApJ...701..508K,2012ApJ...748..142D}) and detailed studies of AGN spectral
energy distributions (SEDs; $\nu F_\nu$) in the near/far-IR (e.g., \citealt{2005AJ....129.1183R,2006ApJS..166..470R,2007ApJ...660..167D,2010ApJ...713..970A,2010ApJ...709..572K,2010ApJ...717.1181P,2014ApJ...790...54C}).
These studies have expanded further based on the all-sky 
{\it Wide-field Infrared Survey Explorer} ({\it WISE}; \citealt{2010AJ....140.1868W}) mid-IR survey data
(e.g., \citealt{2012ApJ...754...45I,2012ApJ...753...30S,2012ApJS..200...17T,2013ApJ...772...26A,2013AJ....145...55Y,2013MNRAS.434..941M,2014ApJ...784..113S,2015ApJ...804...27A}).

There is another dimension in which {\it Spitzer} can significantly expand its studies of
AGN demographics and physics, namely the time domain. 
AGNs are variable sources at all wavelengths 
(e.g., \citealt{2001ASPC..224..265F,2004ApJ...601..692V,2008A&A...486..411S,2009ApJ...707..727A,2010ApJ...721.1014M,2010ApJ...716..530K,2010ApJ...721..715D,2012ApJ...755...60G}).
Their SEDs have a minimum near a rest wavelength of 1~\micron\ and rise towards both the ``blue'' and the ``red'' (e.g., \citealt{1994ApJS...95....1E,2010ApJ...713..970A}). 
The blue, optical side corresponds to emission from the accretion disk surrounding a super massive black hole at the center of an AGN, 
while the red mid-IR side is due to emission by hot dust (e.g., \citealt{1978Natur.272..706S,1993ARA&A..31..473A,1995PASP..107..803U})
as shown by near-IR reverberation mapping studies of nearby AGNs (e.g, \citealt{2006ApJ...639...46S,2014ApJ...788..159K,2014ApJ...788...48S,2015ApJ...801..127V,2015ApJ...814L..12J})
but also by direct observations of the torus (\citealt{2014MNRAS.439.1648A}).
Depending on redshift, the 3.6--4.5~\micron\ emission probes from near the SED minimum ($z>3$) into the dust-dominated regime ($z<1$).

We know a great deal about how quasars vary, but we know relatively little about the origins of that variability.
Possible origins include accretion disk instabilities, surface temperature fluctuations, or variable heating from coronal X-rays 
(e.g., \citealt{1976MNRAS.175..613S,1993A&A...272....8R,1995ApJ...441..354C,2014ApJ...783..105R}). 
The introduction of stochastic process models, particularly the damped random walk 
(DRW; \citealt{2009ApJ...698..895K,2010ApJ...721.1014M,2010ApJ...708..927K,2011AJ....141...93B,2011ApJ...728...26M,2011ApJ...735...80Z,2012ApJ...753..106M,2012ApJ...760...51R,2013ApJ...765..106Z}), has provided a means to quantitatively describe the variability of 
increasingly improving (in length and cadence) quasar light curves from optical surveys (such as SDSS, \citealt{2010ApJ...721.1014M},
OGLE, \citealt{2010ApJ...708..927K}, La Silla-Quest, \citealt{2015ApJ...810..164C}, or CRTS, \citealt{2015MNRAS.453.1562G}) or reverberation mapping campaigns (e.g., \citealt{2010ApJ...721..715D,2012ApJ...755...60G,2015ApJ...806..128D}).
The DRW model is characterized by two parameters, the time scale and amplitude, where the time scale
has been tied to either disk orbital or thermal time scales (\citealt{2009ApJ...698..895K}). While it
well-describes quasar light curves with a typical cadence of days (\citealt{2013ApJ...765..106Z}), there appears to be a break in the power spectrum distribution at short times scales
observed for highly sampled {\it Kepler} AGNs (\citealt{2011ApJ...743L..12M,2015MNRAS.451.4328K}).
Sparser light curves for hundreds or thousands of AGNs are typically analyzed with an ensemble structure function analysis
(e.g., \citealt{2002MNRAS.329...76H,2004ApJ...601..692V,2005AJ....129..615D,2010ApJ...714.1194S}), commonly described by a single power-law
function with the slope index $\gamma$ and amplitude $S_0$ at a fixed time scale, 
although on long time scales there is clearly a break (\citealt{2012ApJ...753..106M}).

In \cite{2010ApJ...716..530K}, we used a structure function analysis to study the 
ensemble quasar variability of the largest repeatedly observed mid-IR AGN sample from the {\it Spitzer} Deep, Wide-Field Survey.
The survey observed the NDWFS Bo{\"o}tes Field ((RA, Decl.$)=($14:30, $+$34:05)) four times between 2004 and 2008 (\citealt{2004ApJS..154..48E,2009ApJ...701..428A}).
It is a unique 9~deg$^2$ area of the sky, far from the Galactic plane ($\sim$67~deg), 
that has been observed at many different wavelengths to provide a good census of extragalactic objects. 
The AGN and Galaxy Evolution Survey (AGES; \citealt{2012ApJS..200....8K}) provides redshifts for over 23,500 objects.
The NOAO Deep Wide-Field Survey (NDWFS; \citealt{1999ASPC..191..111J}) provides deep optical images and photometry in the $B_{\rm w}RI$ bands,
while the {\it Spitzer} Deep, Wide-Field Survey (SDWFS; \citealt{2009ApJ...701..428A}) provides magnitudes in four mid-IR bands at 
3.6, 4.5, 5.8, and 8.0 microns. It was also observed at X-ray (XBo\"otes; \citealt{2005ApJS..161....1M}), at UV by {\it GALEX} (\citealt{2005ApJ...619L...1M})
at 24~\micron\ (MAGES survey), at far-IR by {\it Herschel Space Observatory} (\citealt{2010A&A...518L...1P}), 
and radio (FIRST; \citealt{1995ApJ...450..559B} and WSRT; \citealt{2002AJ....123.1784D}) wavelengths.

In this paper, we extend our previous analysis using the Decadal IRAC Bo{\"o}tes Survey (DIBS), which stretched
the time baseline of the survey to 10 years by adding a fifth epoch. This expansion enables two improvements over our earlier study.
First, the longer temporal baseline provided by the fifth epoch significantly improves the structure function slope measurement, 
our ability to characterize the differences between optical and mid-IR variability, and to subdivide the sample based on
other criteria (e.g., X-ray or radio flux). The ensemble quasar variability in mid-IR can be described by a single power-law structure function with the index $\gamma\approx0.5$,
hence increasing the time baseline from 4 (SDWFS) to 10 years (DIBS), typically increases quasar variability by 50\%.
Second, existing mid-IR AGN samples are strongly biased against low luminosity AGN and variability provides a means
of removing this bias to obtain a more complete census -- the addition of the fifth epoch should improve the completeness and/or purity of the
variability selection. Please note, however, that the typical $rms$ of mid-IR quasar variability is $\sim$0.1~mag, which is comparable to the uncertainties provided by DIBS or SDWFS for typical sources, hence we detect only significantly variable or brighter quasars.

The paper is organized as follows. In Section~\ref{sec:data}, we present the available data and data analysis. 
In Section~\ref{sec:definitions}, we introduce the variability measures,
while in Section~\ref{sec:results}, we discuss the results of the quasar selection methods. The structure function analysis is presented in Section~\ref{sec:sfunction},
and the paper is summarized in Section~\ref{sec:summary}.


\section{Data Analysis}
\label{sec:data}

\begin{deluxetable*}{lc|ccccc|ccccc}
\tabletypesize{\scriptsize}
\tablecaption{The 3.6 Micron Light Curves for the DIBS Variability Catalog.\label{tab:SDWFSlc1}}
\tablewidth{0pt}
\tablehead{
\multicolumn{2}{c}{} &
\multicolumn{5}{c}{Measurements} &
\multicolumn{5}{c}{Uncertainties} \\
RA &
Dec & 
$\epsilon_1$ &
$\epsilon_2$ &
$\epsilon_3$ &
$\epsilon_4$ &
$\epsilon_5$ &
$\epsilon_1$ &
$\epsilon_2$ &
$\epsilon_3$ &
$\epsilon_4$ &
$\epsilon_5$ \\
(deg) &
(deg) &
(mag) &
(mag) &
(mag) &
(mag) &
(mag) &
(mag) &
(mag) &
(mag) &
(mag) &
(mag) \\
}
\startdata
217.659222 & 32.452357 & 18.677 & 18.742 & 18.777 & 18.920 & 18.977 & 0.141 & 0.139 & 0.137 & 0.131 &  0.153 \\
217.730372 & 32.452616 & 19.048 & 18.793 & 19.001 & 19.217 & 19.029 & 0.158 & 0.159 & 0.155 & 0.147 &  0.173 \\
217.787424 & 32.452656 & 19.349 & 19.573 & 20.245 & 19.596 & 99.999 & 0.207 & 0.204 & 0.196 & 0.192 & 99.999 \\
217.628739 & 32.452706 & 19.119 & 19.483 & 19.180 & 19.391 & 19.950 & 0.179 & 0.175 & 0.174 & 0.166 &  0.190 \\
217.618530 & 32.452833 & 18.907 & 18.853 & 19.024 & 18.676 & 18.973 & 0.147 & 0.146 & 0.143 & 0.139 &  0.160
\enddata
\tablecomments{$\epsilon_i$ with $i=1,...,5$ are the epoch numbers. The error code for magnitudes, reflecting no measurement, is 99.999.\\
(This table is available in its entirety in a machine-readable form in
the online journal. A portion is shown here for guidance regarding its
form and content.)}
\end{deluxetable*}

\begin{deluxetable*}{lc|ccccc|ccccc}
\tabletypesize{\scriptsize}
\tablecaption{The 4.5 Micron Light Curves for the DIBS Variability Catalog.\label{tab:SDWFSlc2}}
\tablewidth{0pt}
\tablehead{
\multicolumn{2}{c}{} &
\multicolumn{5}{c}{Measurements} &
\multicolumn{5}{c}{Uncertainties} \\
RA &
Dec & 
$\epsilon_1$ &
$\epsilon_2$ &
$\epsilon_3$ &
$\epsilon_4$ &
$\epsilon_5$ &
$\epsilon_1$ &
$\epsilon_2$ &
$\epsilon_3$ &
$\epsilon_4$ &
$\epsilon_5$ \\
(deg) &
(deg) &
(mag) &
(mag) &
(mag) &
(mag) &
(mag) &
(mag) &
(mag) &
(mag) &
(mag) &
(mag) \\
}
\startdata
217.659222 & 32.452357 & 18.875 & 19.046 & 19.653 & 18.559 & 18.317 &  0.192 &  0.158 &  0.170 & 0.172 & 0.164 \\
217.730372 & 32.452616 & 18.844 & 18.433 & 19.255 & 18.652 & 18.526 &  0.170 &  0.143 &  0.152 & 0.150 & 0.143 \\
217.787424 & 32.452656 & 99.999 & 99.999 & 99.999 & 19.855 & 19.689 & 99.999 & 99.999 & 99.999 & 0.321 & 0.304 \\
217.628739 & 32.452706 & 18.837 & 18.673 & 19.173 & 19.397 & 19.468 &  0.206 &  0.171 &  0.184 & 0.177 & 0.167 \\
217.618530 & 32.452833 & 19.314 & 19.029 & 18.570 & 18.273 & 18.591 &  0.173 &  0.143 &  0.161 & 0.158 & 0.147
\enddata
\tablecomments{$\epsilon_i$ with $i=1,...,5$ are the epoch numbers. The error code for magnitudes, reflecting no measurement, is 99.999.\\
(This table is available in its entirety in a machine-readable form in
the online journal. A portion is shown here for guidance regarding its
form and content.)}
\end{deluxetable*}

\begin{figure*}
\centering
\includegraphics[width=7.2cm]{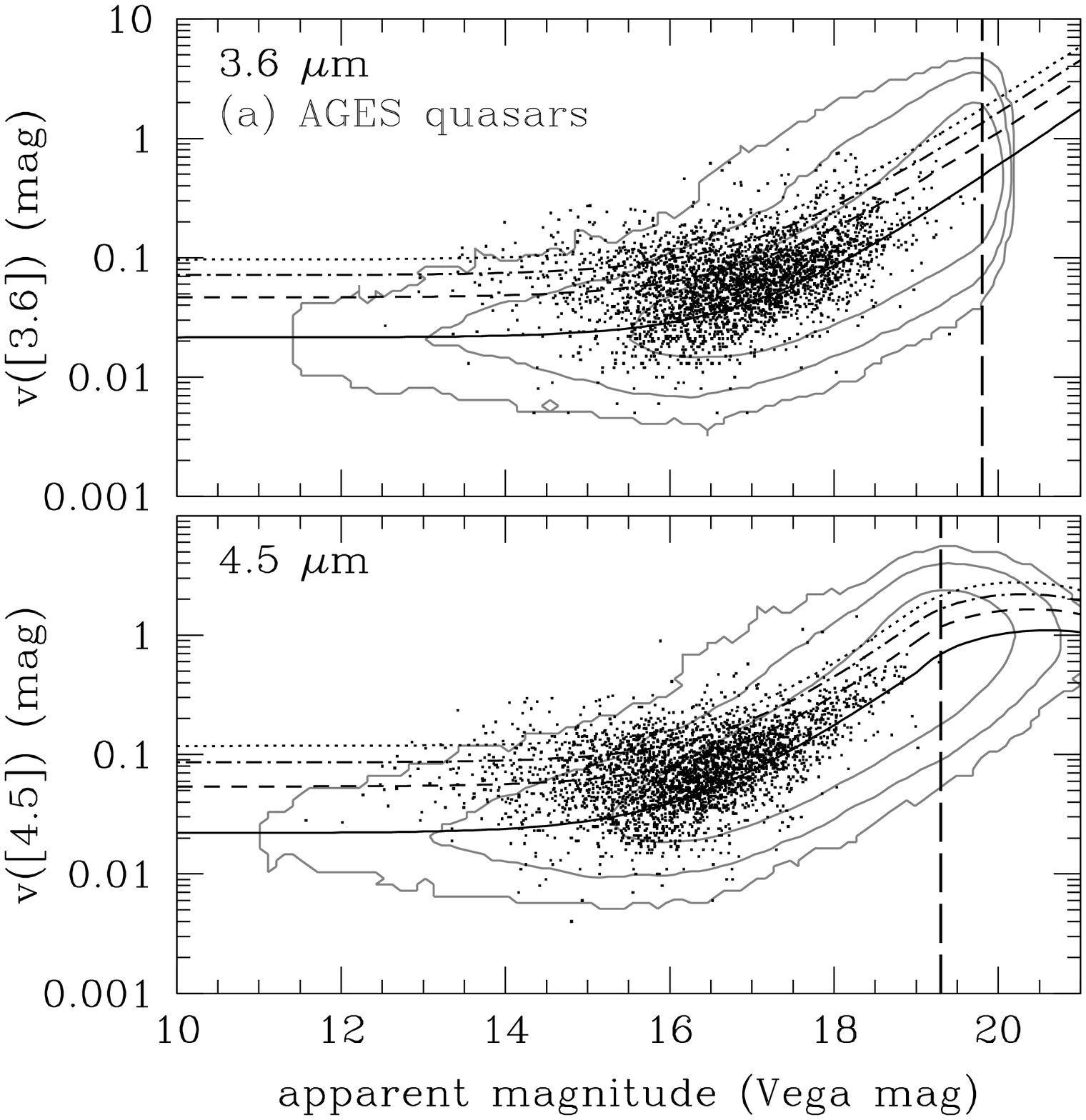}\hspace{0.3cm}
\includegraphics[width=7.2cm]{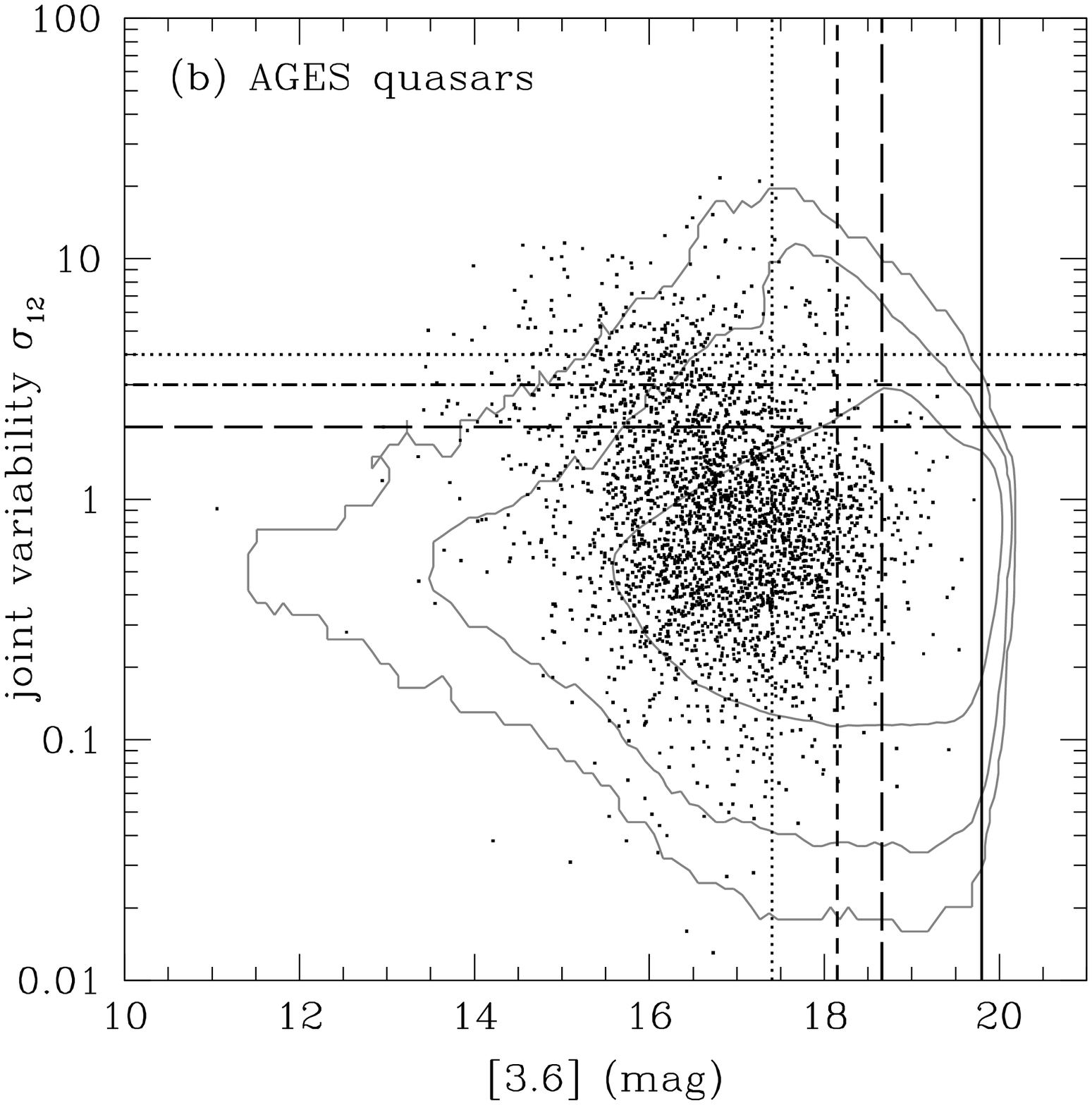}\\
\vspace{0.3cm}
\includegraphics[width=7.2cm]{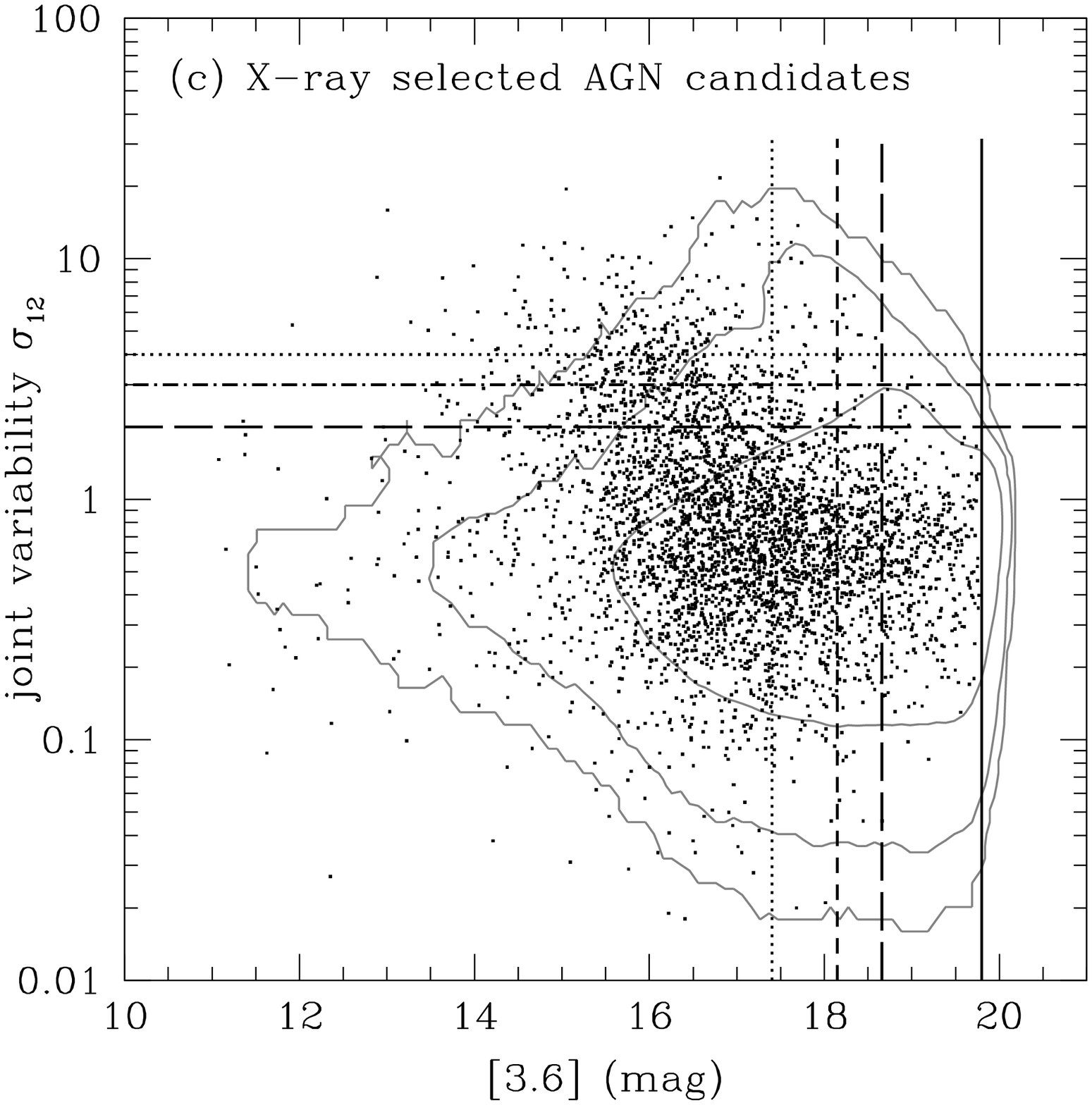}\hspace{0.3cm}
\includegraphics[width=7.2cm]{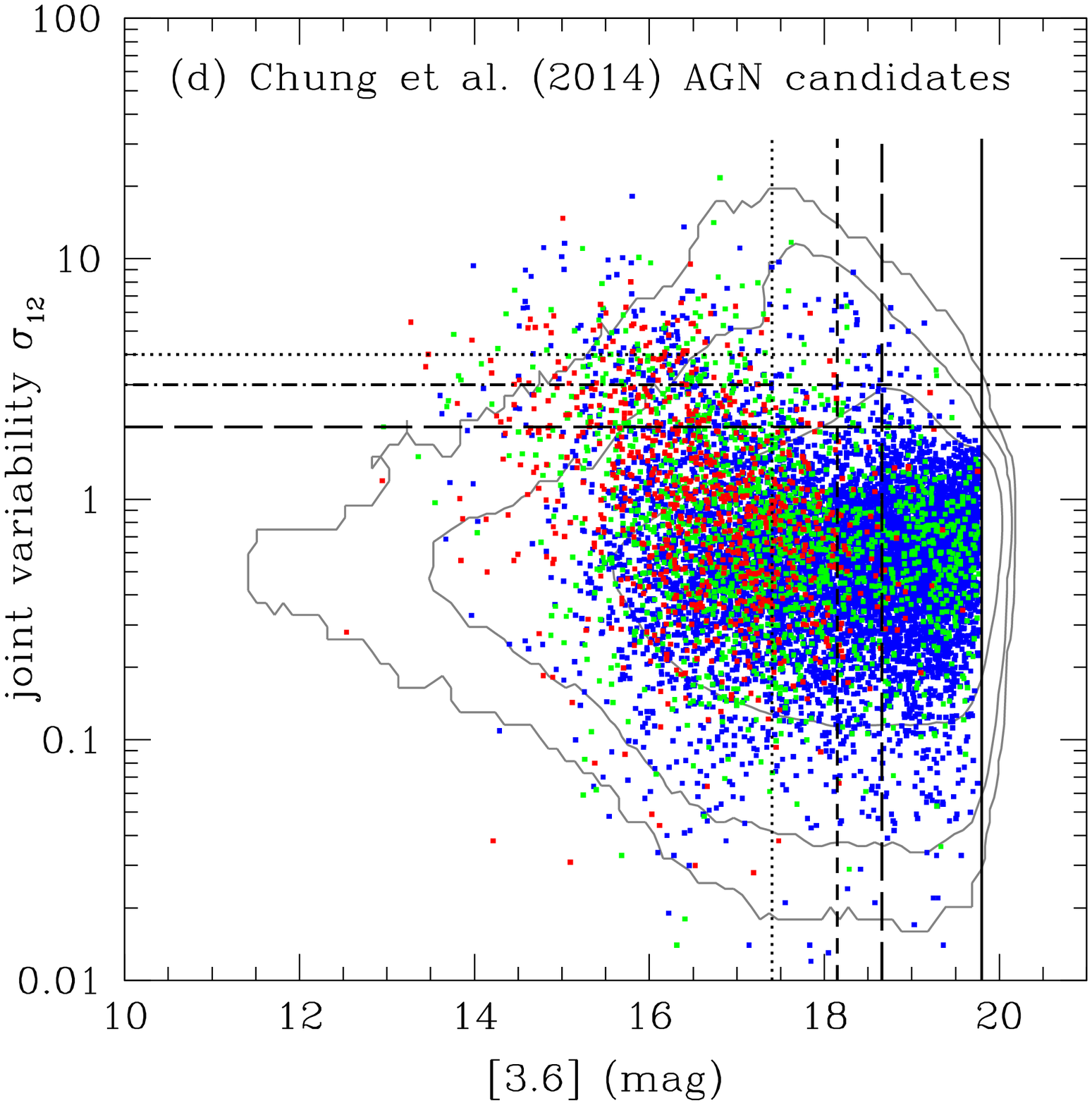}\\
\caption{{\it Panel (a):} Variability in the 3.6 \micron\ and 4.5 \micron\ bands as a function of magnitude for 
all objects with 4 or 5 epochs of data shown as contours. The objects are binned into 0.1 mag and 0.05 log($v$) bins. 
The contours are drawn for 2, 10 and 100 objects per bin, counting from the outer contour. The lower solid line 
shows the median $v_m$ calculated in 0.1 mag bins, and the upper three dashed, dash-dotted and dotted lines show 
variability significance $(v-v_m)/\sigma=1$, 2 and 3  (from bottom to top) with respect to this median. We limit our Photometry Group 2 
and 3 samples to objects with $[3.6] < 19.8$ mag, and $[4.5] < 19.3$ mag (vertical dashed lines). 
{\it Panels (b--d):} Joint variability $\sigma_{12}$ as a function of [3.6] magnitude. Contours have the same 
meaning as in panel (a). The three horizontal lines mark the Variability Levels 2, 3, and 4, from bottom to top, while
the vertical lines mark the mean variability dispersion at 0.05, 0.10 and 0.15 mag (from left to right) and the limiting magnitude
(the rightmost solid line). In the three panels we show
(b) the AGES confirmed quasars, (c) X-ray detected sources, and (d) AGN candidates from \cite{2014ApJ...790...54C}.
Colors in panel (d) denote the \cite{2014ApJ...790...54C} $F$-ratio ranges, where $10<F<30$ is blue, $30<F<80$ is green, and $F>80$ is marked with red (see text). 
\label{fig:var3645}}
\end{figure*}

A detailed description of SDWFS is presented in \cite{2009ApJ...701..428A}. The first epoch was taken on  2004 January 10-14 as 
the IRAC Shallow Survey (\citealt{2004ApJS..154..48E}). Then, we added three more epochs on 2007 August 8-13, 2008 February 2-6, and 2008 March 6-10
as the {\it Spitzer} Deep, Wide-Field Survey (\citealt{2009ApJ...701..428A}; D. Stern PI, PID 40839). The new, fifth epoch was taken on 2014 April 23-29 as the
Decadal IRAC Bo{\"o}tes Survey (DIBS; M.L.N. Ashby PI, PID 10088). The intervals between any two epochs span the range from 1 month to 10 years.

Each epoch consists of $\sim$20,000 individual $256\times256$ pixel IRAC (\citealt{2004ApJS..154...10F}) frames that were combined into a single mosaic. We have 
re-reduced the original SDWFS images and created new mosaics, each approximately $19000\times23000$ pixels, with a pixel scale of 0.60 arcsec.
These mosaics cover the same area of the sky, but at a different pixel resolution from SDWFS ($13600\times16700$ pixels and 0.84~arcsec).
We used the same photometric catalogs as provided in \cite{2009ApJ...701..428A}. Due to the slightly different mosaicing of the fifth epoch
and a slightly modified image subtraction procedure (described below), we report the photometric measurements for a total of 514975 objects.

The photometry of our sources was done using difference imaging methods (e.g., \citealt{1998ApJ...503..325A}), where we use the implementation of \cite{2000AcA....50..421W}.
First, all ten mosaics (five for each of the $3.6\mu$m and $4.5\mu$m bands) were aligned to the same reference frame. We then cut them into a slightly overlapping 
$5\times6$ grid of $3940\times3940$ pixel subimages (with 20 pixel overlaps). Because the {\it Spitzer} point spread function (PSF) is ``triangular'' with diffraction spikes, we pre-filtered
these images to obtain Gaussian-like PSFs (see \citealt{2010ApJ...716..530K} for details), which greatly helps the difference image analysis (DIA) software obtain clean subtractions and photometry. 
Having the corrected images, we chopped each of them into even smaller $2\times2$ grids of $1990\times1990$ pixel chunks (again with 20 pixels of overlap) to speed up the DIA analysis. 
We stacked all five epochs in order to create template images and then use DIA to match and subtract the reference images from each individual epoch.

We measured the fluxes of the objects in the SDWFS catalogs on the template image to photometrically align them to the SDWFS catalog. 
They were measured using the PSF fitting routines from the DIA package of \cite{2000AcA....50..421W}. DIA typically returns underestimated 
uncertainties, hence they were corrected with $\sigma_{\rm new}=\sqrt{(\zeta\times\sigma_{\rm DIA})^2+\epsilon^2}$, where $\zeta=5.1$ and $\epsilon=0.00324$
were set so the new error bars $\sigma_{\rm new}$ match the dispersions for non-variable field objects.
After subtraction of the template, the difference fluxes at the positions of all sources were measured and added to the fluxes measured on the template, 
thereby creating the light curves (Tables~\ref{tab:SDWFSlc1} and \ref{tab:SDWFSlc2}).  
Our catalogs are limited to the objects detected on the template image. A transient which does not produce 
a detectable source in the (averaged) template image and is not present in the SDWFS catalog will not be identified.
The brightest objects create multiple variable artifacts in their vicinity. We masked the regions around these objects using 
the Two Micron All Sky Survey (2MASS) catalog (\citealt{2003tmc..book.....C}), where the masking region radius was $260(1-K/11)$ arcsec for $K<11$~mag stars.

The absolute magnitudes were derived using a standard $\Lambda$CDM cosmological model with
$(H_0, \Omega_M, \Omega_{\rm vac}, \Omega_k)=(70~{\rm{km~s^{-1} Mpc^{-1}}}, 0.3, 0.7, 0.0)$
and using K-corrections derived from \cite{2010ApJ...713..970A}.


\section{Variability}
\label{sec:definitions}

\begin{figure}
\centering
\includegraphics[width=8.2cm]{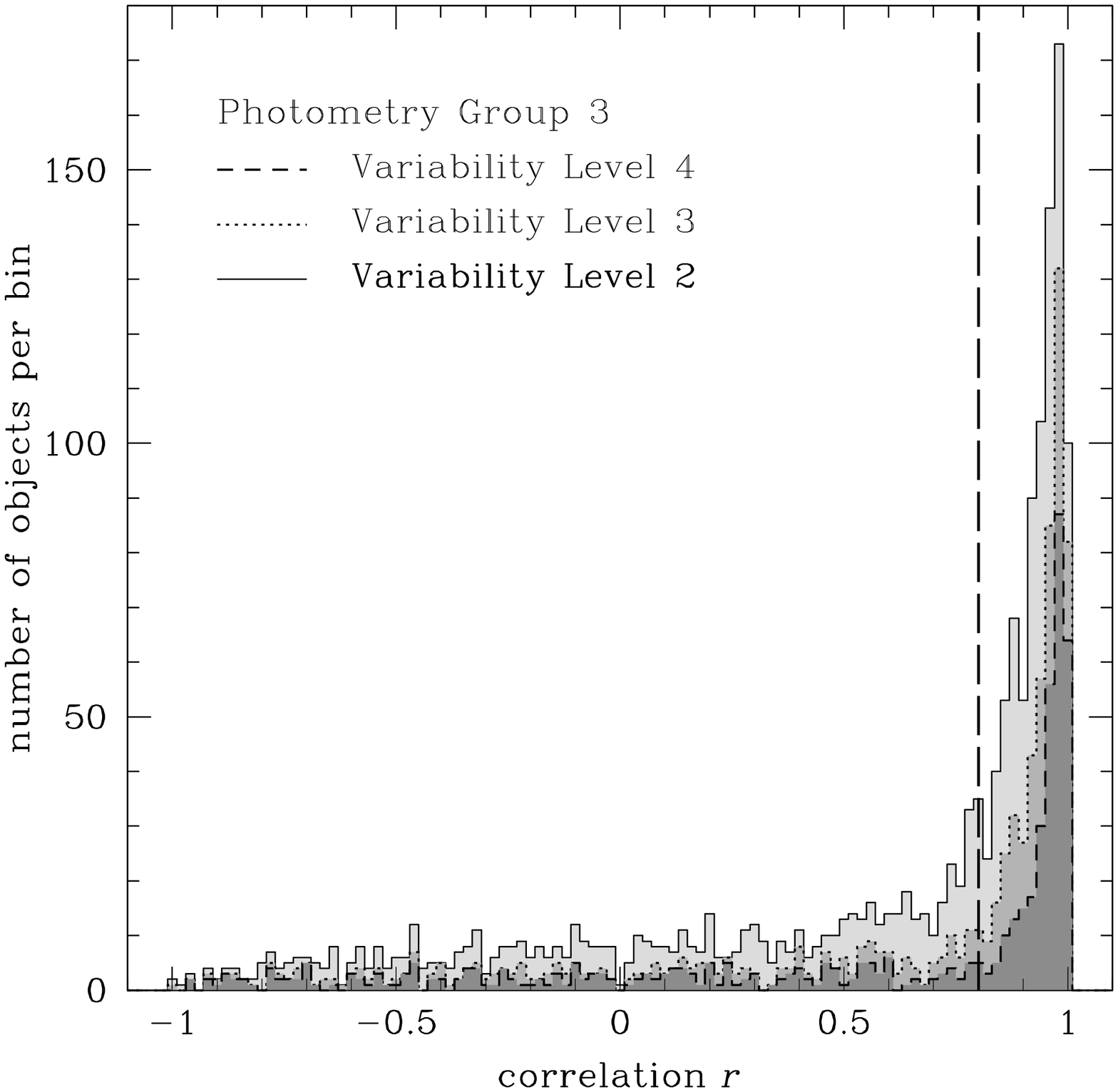}
\caption{Correlation between the 3.6 \micron~and 4.5 \micron~light curves for sources with measured $[5.8]-[8.0]$ colors
(``Photometry Group 3'' sources; see \S\ref{sec:definitions}) with 4 or 5 epochs of data for
Variability Level 2, 3 and 4 ($\sigma_{12}>2$, $>3$ and $>4$) sources.  The true variables are the sources to the right 
of the vertical dashed line, in the high correlation region ($r>0.8$). The poorly correlated ($r<0.5$) 
sources were used to estimate our false positive rates inside the high correlation region (Tables~\ref{tab:resultsDIBS}, \ref{tab:resultsWave}, and \ref{tab:resultsAGES}).
\label{fig:covariance}}
\end{figure}

\begin{figure}
\centering
\includegraphics[width=8.2cm]{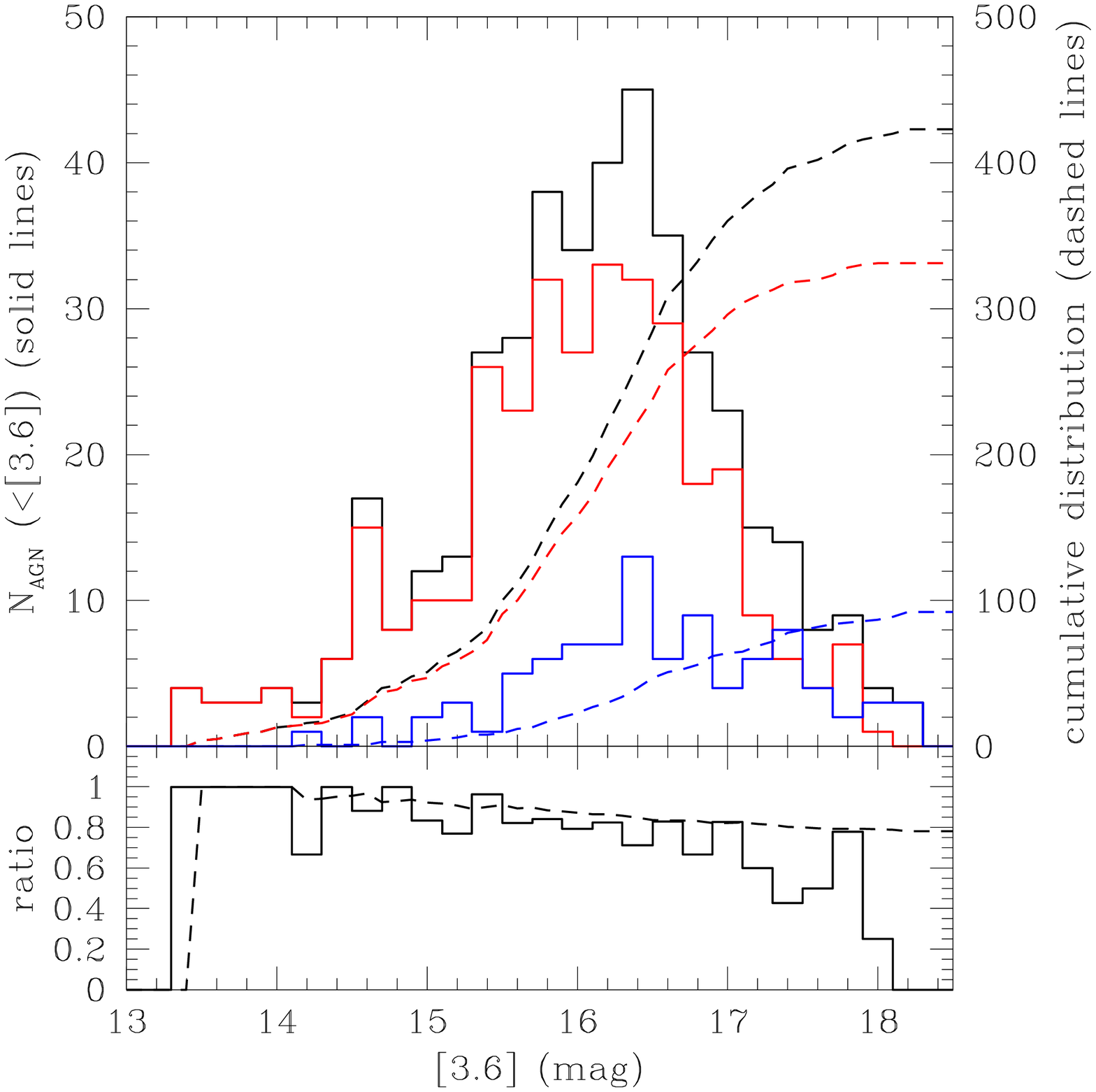}
\caption{{\it Top panel:} Histograms (solid lines, left scale) and cumulative distributions (dashed lines, right scale) for spectroscopically confirmed AGES
quasars with data from all five epochs and  variability exceeding VL2 as a function of observed [3.6] magnitude. 
Red, blue, and black lines are for high correlation sources with $r>0.8$,
for $r<0.8$, and all AGNs, respectively.
{\it Bottom panel:} Ratios of histograms and cumulative distributions for highly correlated AGNs to all AGNs. 
\label{fig:corr_hist}}
\end{figure}

We were interested in sources with at least three epochs ($N_{\rm ep}\gtrsim3$) measured in both the 3.6~\micron\ and 4.5~\micron\ bands. From the fluxes we compute the 
standard deviations $v([X])_i$ as a function of apparent magnitude $m([X])_i$ (Figure~\ref{fig:var3645}, panel (a)).
We also calculated the variability covariance $C_{12}^i$ and Pearson's 
correlation $r^i$ coefficients between the two bands, where
\begin{eqnarray}
v^2([X])^i & = & {\frac{1}{N_{\rm ep}-1}\sum_{j=1}^{N_{\rm ep}}\left(m([X])^i_j-\langle m([X])^i \rangle\right)^2} ,\\
C_{12}^i & = & \frac{1}{N_{\rm ep}-1}\sum_{j=1}^{N_{\rm ep}}\left(m([3.6])^i_j-\langle m([3.6])^i \rangle\right) \nonumber \\
& \times & \left(m([4.5])^i_j-\langle m([4.5])^i \rangle\right), \\
r^i & = & \frac{C_{12}^i}{v([3.6])^i v([4.5])^i},
\end{eqnarray}
and $\langle m([X])^i \rangle$ is the average magnitude for the $i$th light curve in band $X$.
A correlation coefficient of $r=1$ ($-1$, $0$) means that
the 3.6~\micron\ and 4.5~\micron\ light curves are perfectly correlated (anti-correlated, uncorrelated).  We expect the majority of sources in the field to be non-variable galaxies, 
hence the distribution of $v([X])_i$ is dominated by contributions from noise and not true variability. 
In order to select truly variable objects, we defined a variability significance threshold. Let $v_m([X])$ be the median dispersion as a function
of magnitude $m_i$ and $\sigma([X])$ be the dispersion of the $v([X])_i$ values around the median. 
The more distant the source is from the median, $(v([X])_i-v_m([X]))/\sigma([X])>0$, and the more its
3.6~\micron\ and 4.5~\micron\ light curves are correlated, the more likely it is to be the true variable.
The typical $rms$ of mid-IR quasar variability (marked by dots in Fig.~\ref{fig:var3645}) is $\sim$0.1~mag, which is comparable to the uncertainties
provided by DIBS or SDWFS, hence we can detect only bright and/or significantly variable quasars.
We define a combined significance of the variances in the light curves as 

\begin{eqnarray}
\sigma_{12}^2(i) & = &  \left(\frac{v([3.6])_i-v_m([3.6])(m_i)}{\sigma([3.6])(m_i)}\right)^2 + \nonumber  \\
              & + &   \left(\frac{v([4.5])_i-v_m([4.5])(m_i)}{\sigma([4.5])(m_i)}\right)^2 ,
\end{eqnarray}
which quantifies the degree to which the object deviates from the median variances in both bands. The 
best candidates for true variables are objects with
positive excess variances (large $\sigma_{12}$ and $v>v_m$) and strong correlations $r\simeq 1$ between the two bands.
We define three Variability Levels 2, 3, and 4 (in short VL2, VL3, and VL4) for sources exceeding the joint variability significance
$\sigma_{12}>2$, 3, and 4, respectively, that will be frequently used to select variable objects and discuss selection yields (see Figure~\ref{fig:var3645}, panels (b-d)). By definition half of the sources have negative $(v([X])_i-v_m([X]))/\sigma([X])$ and anomalously non-variable sources could also
give rise to high values of $\sigma_{12}$. The distributions of $(v([X])_i-v_m([X]))$ are, however, very non-symmetric and while for the 3.6~\micron\ or 4.5~\micron\
bands there are 7443 or 7538 objects with $(v([X])_i-v_m([X]))/\sigma([X])>1$ and $\sigma_{12}(i)>2$, there are only 0 or 17
sources with $(v([X])_i-v_m([X]))/\sigma([X])<-1$ and $\sigma_{12}(i)>2$. This means that the high variability significance sources are
those with positive (high) variability.

\begin{figure}
\centering
\includegraphics[width=8.2cm]{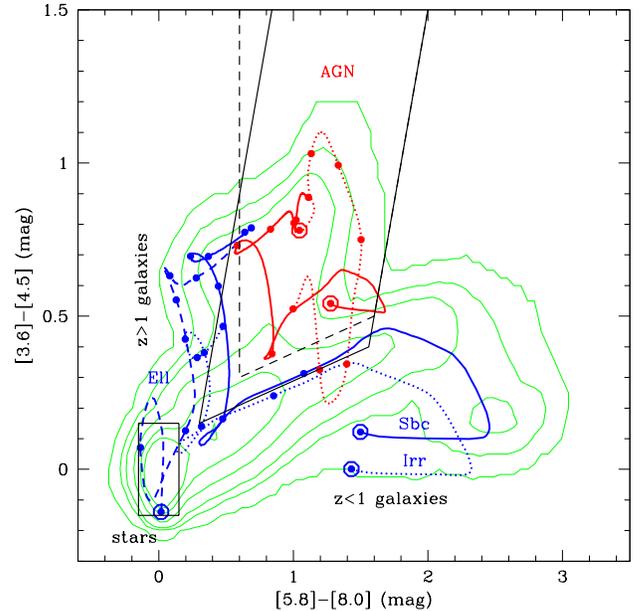}
\caption{Mid-IR color-color distribution of the Photometry Group 3 sources in DIBS (green). 
The contours are based on 39591 objects that are counted in 0.05 mag bins in both axes.
The contours are at levels of 2, 10, 20, 50, 100, and 150 (counting from the outside).
In order to understand color-color selections, we plot the \cite{2010ApJ...713..970A} color-color tracks as a function of redshift 
for common object classes such as AGNs (red dotted line), spiral (Sbc; solid blue line), elliptical (Ell; dashed blue line), and irregular (Irr; dotted blue line) galaxies.
We also show a combined light of an AGN and its host spiral galaxy, contributing light equally (red solid line). The dots on each track mark redshift increment by 0.5 (1.0)
from $z=0$ to 3 (to 6) for galaxies (AGNs) that start at $z=0$ bulls eye. The solid black trapezoid is the modified AGN wedge, the dashed wedge is the old AGN wedge, and the 
black solid rectangle is the stellar box.
\label{fig:col-col_explain}}
\end{figure}

The distribution of correlation strengths $r$ for sources with $\sigma_{12}$ exceeding a range of variability significance thresholds is shown in Figure~\ref{fig:covariance}. 
It is obvious that the true variables tend to be objects with high correlations $r>0.8$ (dashed vertical line).
We can estimate the contamination as follows. We assume that uncorrelated sources ($r<0.5$) in Figure~\ref{fig:covariance} represent the density of
false positives. If the number of identified variables with $\sigma_{12} > 2$ (or 3, 4) and $r>0.8$ is
$N_{r>0.8}$, and the number of sources with 
$\sigma_{12} > 2$ (or 3, 4) and $r<0.5$ is $N_{r<0.5}$, then we estimate that the number of false positives is
$F=(0.2/1.5)N_{r<0.5}$, where $0.2/1.5$ is the ratio of the intervals $r>0.8$ and $-1 \le r \le 0.5$.
The fraction of the identified variables that are false positives is then $F/N_{r>0.8}$. 
In Figure~\ref{fig:corr_hist}, we show both the histograms and cumulative distributions
for spectroscopically confirmed quasars from the AGES survey with the variability exceeding VL2 as a function of the
observed [3.6] magnitude. Bright sources with highly correlated light curves ($r>0.8$) constitute $\sim$100\% of the AGN sample,
but this proportion decreases to $\sim$80\% with decreasing source brightness.
The variability properties of DIBS sources are presented in Table~\ref{tab:varcat}.

Knowledge of the mid-IR colors is key to assigning objects to various classes such as AGNs, galaxies or stars (Figure~\ref{fig:col-col_explain}). 
We use the same definitions as in \cite{2010ApJ...716..530K}, where the stellar box is $-0.15<[3.6]-[4.5]<0.15$ and $-0.15<[5.8]-[8.0]<0.15$,
the old AGN wedge (\citealt{2005ApJ...631..163S}) is defined as
$[5.8]-[8.0]>0.6$~mag (left side),\\
$[3.6]-[4.5]>2.5\times([5.8]-[8.0])-3.5$~mag (right side),\\
$[3.6]-[4.5]>0.2\times([5.8]-[8.0])+0.18$~mag (bottom),\\ and
the modified AGN wedge is defined as \\
$[3.6]-[4.5]<2.5\times([5.8]-[8.0])-0.6$~mag (left side),\\
$[3.6]-[4.5]>2.5\times([5.8]-[8.0])-3.5$~mag (right side),\\
$[3.6]-[4.5]>0.2\times([5.8]-[8.0])+0.08$~mag (bottom).\\ 
The mid-IR colors depend vitally on the brightness of objects, therefore we divide our sources into three classes 
that increase with our ability to characterize sources in the mid-IR color-color plane: 

\begin{itemize}  
\item Photometry Group 1 (PG1) consists of all objects that after masking have 4 or 5 epochs of data, with no constraints on either 
    their magnitudes or magnitude uncertainties.  There are 438270 Photometry Group 1 objects after masking, of which 
    9811, 4229 and 2472 meet the Variability Level 2, 3 and 4 criteria.

\item Photometry Group 2 (PG2) is the subset of the Photometry Group 1 sources for which we can measure the [3.6]$-$[4.5] color.  These
   sources must satisfy $[3.6] < 19.8$ mag and $[4.5] < 19.3$ mag with uncertainties in these
   bands smaller than 0.1~mag.  There are 214433 Photometry Group 2 objects, of which 5325, 2878 and 1851 meet
   the Variability Level 2, 3 and 4 criteria.

\item Photometry Group 3 (PG3) is the subset of Photometry Group 2 objects for which we can also measure the [5.8]$-$[8.0] color.  These
   sources are required to have uncertainties smaller than $<0.2$ mag for the 5.8 and 8.0 \micron~bands.  There
   are 39591 Photometry Group 3 objects, of which 1615, 840 and 504 meet the Variability Level 2, 3 and 4 criteria.
\end{itemize}
In Table~\ref{tab:resultsDIBS}, we present the significance statistics for the DIBS sources 
as a function of Photometry Group, Variability Level, and the mid-IR color-color class.
The most variable, highly correlated ($r>0.8$) sources in the field are AGNs and stars, peaking at $\sigma_{12}\approx 50$.

\begin{deluxetable*}{lcccccccrrrrrrc}
\centering
\tabletypesize{\scriptsize}
\tablecaption{DIBS Variability Catalog\label{tab:varcat}}
\tablewidth{0pt}
\tablehead{
Object Name &
RA &
Dec & 
[3.6] &
[4.5] &
[5.8] &
[8.0] &
N$\epsilon$&
$v_{[3.6]}$&
$v_{[4.5]}$&
$C_{12}$ &
r &
$\sigma_{12}$ & 
M$^{\star}$ \\
DIBS & (deg) & (deg) & (mag) & (mag) & (mag) & (mag)  &   & (mag) & (mag) & (mag$^2$) &  &  & \\ 
\hline
}
\startdata
J143038.21+322708.5 & 217.6592 & 32.4523 & 18.56 & 18.56 & 99.99 & 17.03 & 5 &   0.13 &   0.51 &  $-$0.04 &  $-$0.65 &   0.85 & 1 \\
J143055.29+322709.4 & 217.7303 & 32.4526 & 18.89 & 18.51 & 17.44 & 16.26 & 5 &   0.15 &   0.33 &     0.01 &     0.22 &   0.26 & 1 \\
J143108.98+322709.6 & 217.7874 & 32.4526 & 19.52 & 20.45 & 99.99 & 99.99 & 1 &  99.99 &  99.99 &    99.99 &    99.99 &   1.39 & 1 \\
J143030.90+322709.7 & 217.6287 & 32.4527 & 19.31 & 18.51 & 99.99 & 99.99 & 5 &   0.33 &   0.35 &     0.06 &     0.48 &   0.28 & 1 \\
J143028.45+322710.2 & 217.6185 & 32.4528 & 18.81 & 18.50 & 99.99 & 99.99 & 5 &   0.13 &   0.41 &     0.02 &     0.29 &   0.59 & 1
\enddata
\tablecomments{Right ascension and declination (magnitudes and variability measures) are provided to four (two) decimal places in order to fit the table into the page, 
while in the electronic version they are in their full form.  The Vega magnitudes presented in the table are measured in 4 arcsec apertures (see \citealt{2009ApJ...701..428A}). 
The error code for magnitudes, reflecting no measurement, is 99.999. M$^{\star}$ is the {\sc 2MASS} mask, where 0 (1) means affected (unaffected) by a bright star. N$\epsilon$ is the number of epochs used in variability calculations.\\
(This table is available in its entirety in a machine-readable form in the online journal. A portion is shown here for guidance regarding its form and content.)}
\end{deluxetable*}


\section{Results}
\label{sec:results}

\begin{figure}
\centering
\includegraphics[width=8.2cm]{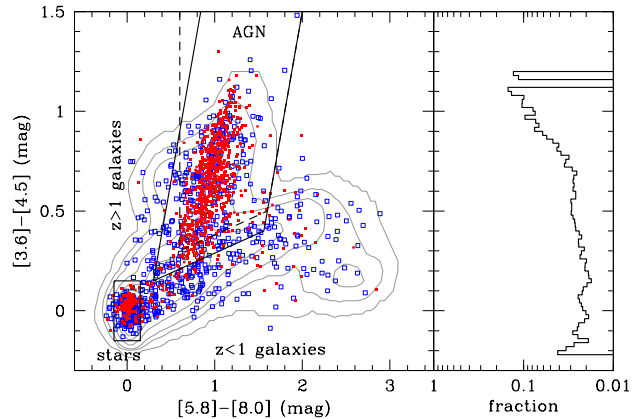}
\caption{
{\it Left panel:} Mid-IR color-color distribution of all Photometry Group 3, Variability Level 2 sources in DIBS. Red points mark highly correlated ($r>0.8$) sources,
while the open blue squares mark less correlated sources ($r<0.8$).
The \cite{2005ApJ...631..163S} AGN selection region is shown by the dashed wedge, the modified wedge from \cite{2010ApJ...716..530K} is shown as
the solid line trapezoid, and the stellar box is marked with a rectangle. The contours are based on 39591 Photometry Group 3 objects, that are counted in 0.05 mag bins in both axes. The contours are at levels of 2, 10, 20, 50, 100, and 150 (counting from the outside). {\it Right panel:} Histogram of the fraction of the 
Photometry Group 2, Variability Level 2 objects relative to all Photometry Group 2 objects as a function of the  $[3.6]-[4.5]$ color.
\label{fig:colcol_hist}}
\end{figure}

\begin{figure*}
\centering
\includegraphics[width=8.2cm]{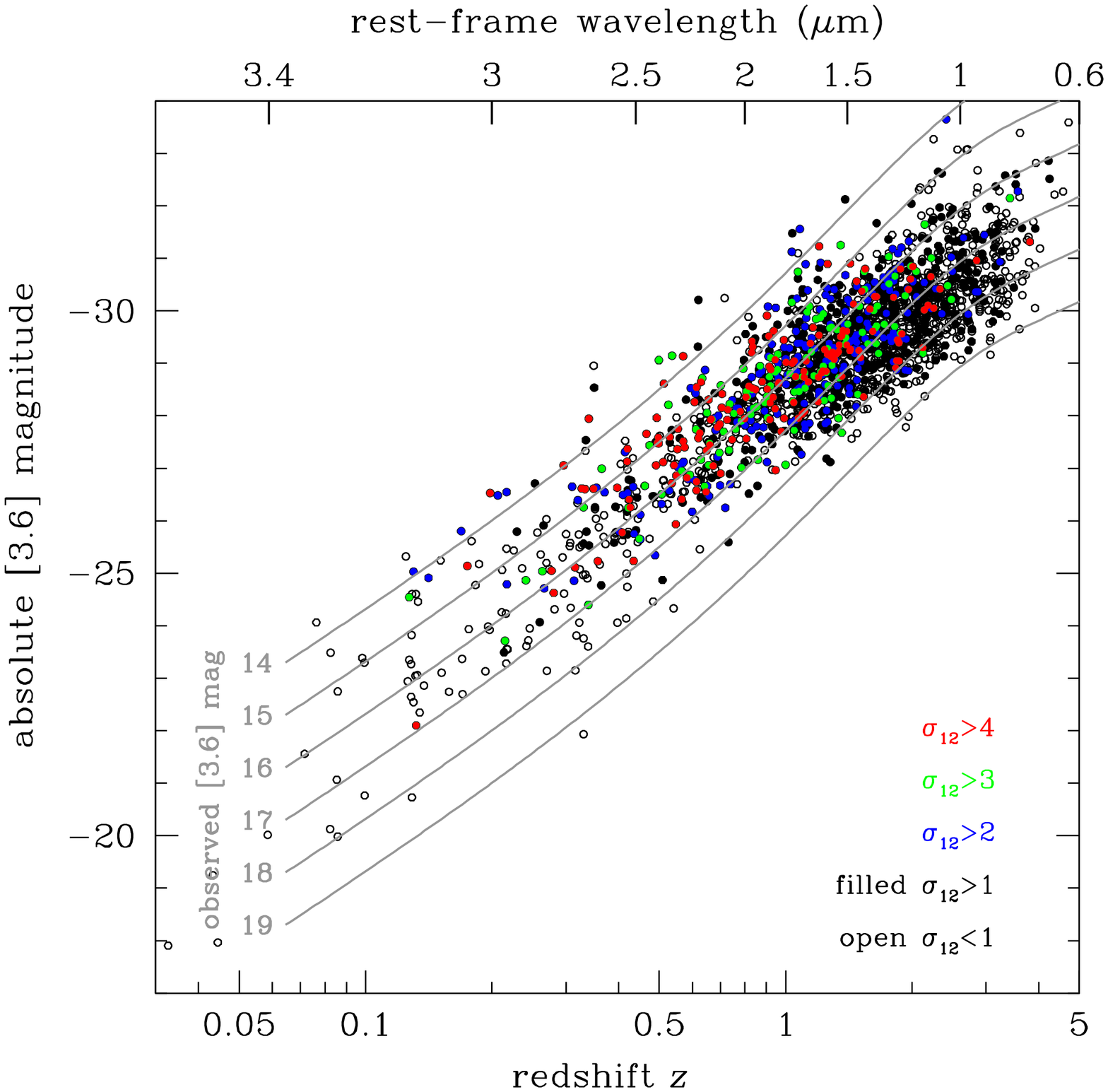}
\hspace{0.3cm}
\includegraphics[width=8.2cm]{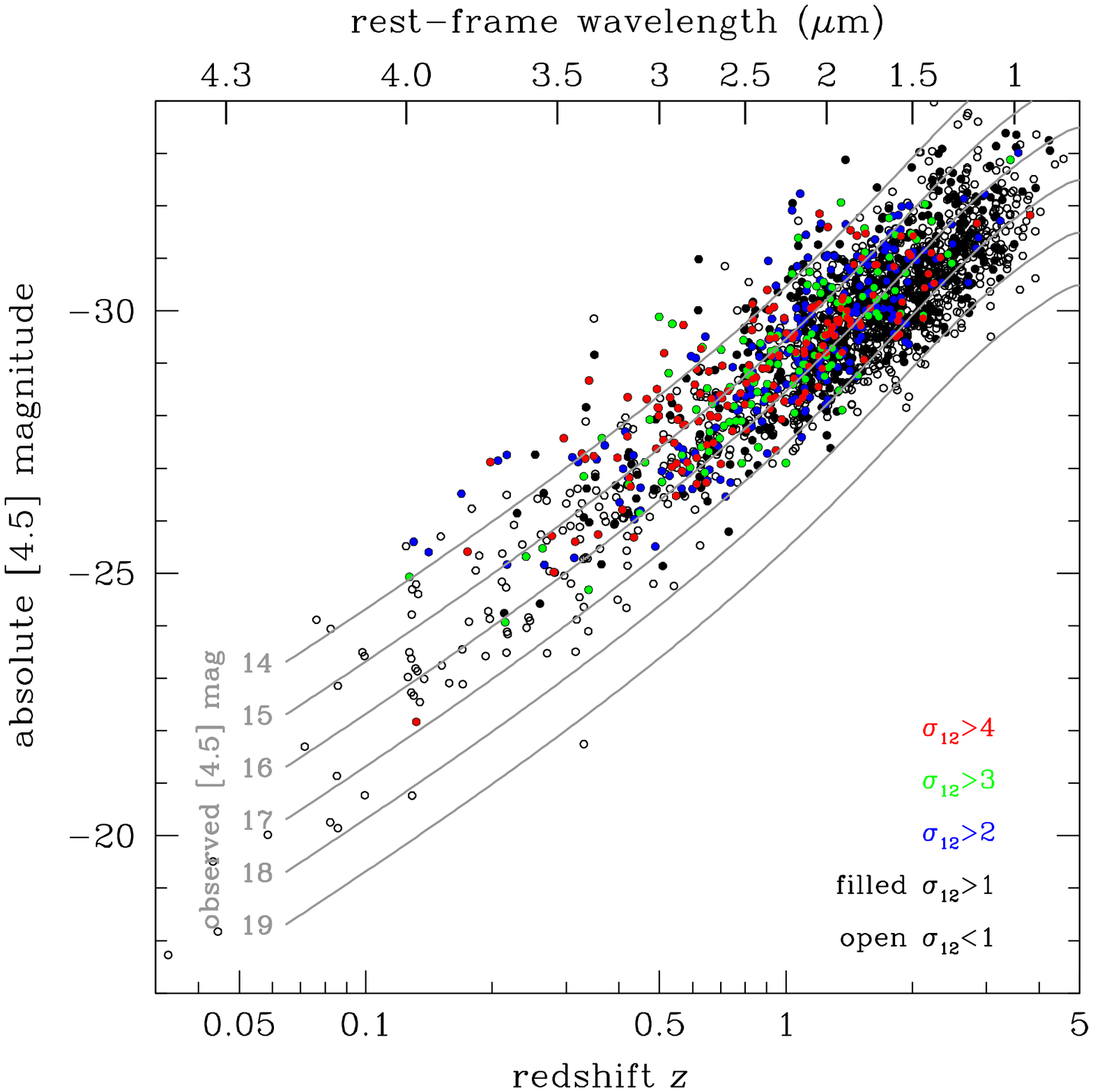}
\caption{Spectroscopically confirmed AGNs from the AGES survey. In the left (right) panel, we show their [3.6] ([4.5]) absolute magnitudes as a 
function of redshift. The points are color coded, where the black open (filled) points mark the variability significance $\sigma_{12}<1$ ($\sigma_{12}>1$),
blue is VL2, green is VL3, and red is VL4.  We also show lines of the constant observed magnitude and the rest-frame wavelength at which the observed light was emitted. 
\label{fig:z_abs}}
\end{figure*}

We will now methodically study the variability of mid-IR sources based on their classifications.

We will start with a general description of all variable objects.
There are 438270 Photometry Group 1 objects (with measurements in 4 or 5 epochs in both bands). There are 9811 (2.2\%) sources that show 
variability $\sigma_{12}>2$ (Variability Level 2) and 1569 (0.4\%) objects that also show a highly significant correlation ($r>0.8$).
For the sources with well-defined $[3.6]-[4.5]$ colors (214433 objects), we find 5325 Variability Level 2 objects (2.5\%) and
1206 (0.6\%) with a high correlation. Photometry Group 3, the sources with well measured $[3.6]-[4.5]$ and $[5.8]-[8.0]$ colors,
contains 39591 sources, of which 1615 (4.1\%) are Variability Level 2 objects and 870 (2.2\%) are also highly correlated.
Increasing the Photometry Group, means focusing on bright sources, where it is easier to measure variability, so it is not surprising
that the variable fraction increases.
For example, for Photometry Group 1, Variability Level 2 and highly correlated ($r>0.8$) sources with $[3.6]<20$, 19, 18, 17, and 16 mag, 
we find  
$1569/438270=0.4\%$,
$1500/334096=0.5\%$,
$1120/175827=0.6\%$,
$833/68330=1.2\%$, and
$487/19132=2.6\%$ variable sources, respectively.

\begin{deluxetable*}{ccccccc}
\tabletypesize{\scriptsize}
\tablecaption{Mid-IR Variable Objects in the Decadal IRAC Bo{\"o}tes Survey.\label{tab:resultsDIBS}}
\tablewidth{0pt}
\tablehead{
\makebox[1.5cm][c]{Variability} & 
\makebox[1.5cm][c]{Variability} & 
\makebox[2.5cm][c]{Old AGN wedge} & 
\makebox[2.5cm][c]{Modified AGN wedge} & 
\makebox[2.0cm][c]{Stellar box} & 
\makebox[2.0cm][c]{Outside$^*$} & 
\makebox[2.0cm][c]{All objects}\\
Level & $\sigma_{12}$ & & & & &  
}
\startdata
\multicolumn{7}{c}{Number (fraction/false positive rate$^{**}$) of variable objects: Photometry Group 1} \\
\cline{1-7}\\
 Total & ---  & 42689            & 74751             & 40343            & 323176               & 438270    \\
2      & $>2$ & 684 (1.6\%/14\%) & 803 (1.1\%/21\%)  & 178 (0.4\%/37\%) & 588 (0.2\%/100\%)    & 1569 (0.4\%/57\%)  \\
3      & $>3$ & 377 (0.9\%/11\%) & 438 (0.6\%/18\%)  & 113 (0.3\%/21\%) & 256 (0.1\%/100\%)    & 807 (0.2\%/47\%)  \\
4      & $>4$ & 213 (0.5\%/13\%) & 242 (0.3\%/20\%)  & 83 (0.2\%/17\%)  & 152 (0.0\%/100\%)    & 477 (0.1\%/48\%)  \\
\\
\multicolumn{7}{c}{Number (fraction/false positive rate$^{**}$) of variable objects: Photometry Group 2} \\
\cline{1-7}\\
 Total & ---  & 29339            & 55055            & 18108             & 141270               & 214433  \\
2      & $>2$ & 651 (2.2\%/9\%)  & 755 (1.4\%/15\%) & 164 (0.9\%/19\%)  & 287 (0.2\%/100\%)    & 1206 (0.6\%/36\%)  \\
3      & $>3$ & 371 (1.3\%/8\%)  & 427 (0.8\%/14\%) & 108 (0.6\%/11\%)  & 167 (0.1\%/100\%)    & 702 (0.3\%/34\%)  \\
4      & $>4$ & 208 (0.7\%/11\%) & 235 (0.4\%/17\%) &  80 (0.4\%/10\%)  & 111 (0.1\%/100\%)    & 426 (0.2\%/38\%)   \\
\\
\multicolumn{7}{c}{Number (fraction/false positive rate$^{**}$) of variable objects: Photometry Group 3} \\
\cline{1-7}\\
 Total & ---  &  6692            & 12801            & 7477              & 19313                & 39591  \\
2      & $>2$ &  593 (8.9\%/2\%) & 652 (5.1\%/3\%)  & 143 (1.9\%/11\%)  & 75 (0.4\%/38\%)      & 870 (2.2\%/7\%)  \\
3      & $>3$ &  341 (5.1\%/1\%) & 376 (2.9\%/3\%)  &  95 (1.3\%/4\%)   & 43 (0.2\%/34\%)      & 514 (1.3\%/5\%)  \\
4      & $>4$ &  192 (2.9\%/2\%) & 209 (1.6\%/3\%)  &  68 (0.9\%/3\%)   & 27 (0.1\%/40\%)      & 304 (0.8\%/6\%) 
\enddata
\tablecomments{$^*$ -- Column ``Outside'' includes objects that are both outside of the ``Modified AGN wedge'' and the ``Stellar box''.\\
$^{**}$ -- The fraction of variable sources is a ratio of the number of variable objects above a given 
level of $\sigma_{12}$ and $r>0.8$ to the total number of objects in a respective region. The false positive rate is 
estimated from the number of sources above the same level of $\sigma_{12}$ but with $r<0.5$ (see \S\ref{sec:definitions} and Figure~\ref{fig:covariance}).}
\end{deluxetable*}

In Figure~\ref{fig:colcol_hist}, we examine the distribution of  variable
Photometry Group 3 sources in the mid-IR color-color plane. 
We only consider this group because both colors are needed to assign classes to objects based on
 their location on the color-color plane. Out of 39591 objects, 870 (2.2\%) are variable
(VL2) with as little as 7\% contamination (Table~\ref{tab:resultsDIBS}). Most of them (652; 75\%) are AGN and are located in the modified AGN wedge. There are also 143
VL2 stars (16\%) and 75 (9\%) objects outside these regions, hence likely  AGNs with bright host galaxies.
Increasing the significance threshold to VL3, we are left with 514 highly correlated variable objects, of which 376 (73\%) are AGNs,
95 (18\%) are stars, and 43 (8\%) are outside these regions. The final and the highest threshold, VL4, selects 304 sources, of which
209 (69\%) are AGNs, 68 (22\%) are stars, and 27 (9\%) are other sources.

It is easy to see that with the increasing significance threshold, the fractional number of AGN decreases while that for stars increases. 
This is likely due to differences in the typical variability amplitudes:
AGNs typical vary by only tenths of a magnitude, while stellar variability 
can be very spectacular, especially in the infrared 
(e.g., Miras or Long Period Variables; see \citealt{2002AJ....123..948S}  for the mid-IR variability and \citealt{2009AcA....59..239S} for the optical variability at 0.8 \micron).
We inspected the mid-IR color-magnitude diagram (CMD; $[3.6]-[5.8]$ color vs. [3.6] magnitude) and found that
(1) all the PG3 VL2 variable objects in the stellar box (but one) are brighter than $[3.6]<15.5$~mag and (2) the stellar locus at the bright end is bi-modal, with a color difference of 0.15 mag, 
and that the variable stars occupy the redder branch.  We suspect that
our typical variable star is an RR Lyrae variable.  These show
mid-IR variability amplitudes of $\sim$0.3~mag (\citealt{2014MNRAS.441..715G}) that are similar to our sources, and with $[3.6]_{\rm abs}\approx-1$~mag
they would have distances (2--20~kpc) consistent with being in the
Galactic halo.  The primary alternative, variable red giants, tend
to have higher amplitudes (0.5--1.0~mag; \citealt{2002AJ....123..948S}) and would have implausibly large distances (100~kpc) because they
typically have $[3.6]_{\rm abs}<-9$~mag (e.g., \citealt{2006AJ....132.2268M}).
Since a fair fraction of variable stars are bright, we also note that some of their variability may not be real but rather be due to residuals from Gaussian filtering the images. 
By increasing the significance threshold, we lose less variable quasars more quickly than the more variable stars. 

\begin{figure}
\centering
\includegraphics[width=8.2cm]{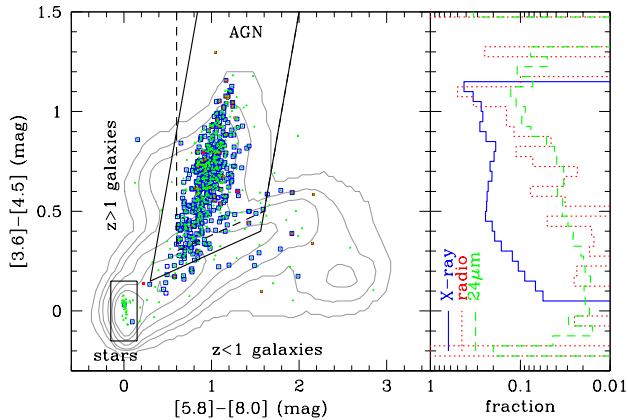}
\caption{
{\it Left panel:} Mid-IR color-color distribution of the Photometry Group 3, Variability Level 2 sources selected as AGNs based on
 their X-ray (blue), radio (red) or 24 \micron~emission (green).  
The contours and selection regions are the same as in Figure~\ref{fig:colcol_hist}.
{\it Right panel:} Histograms of the fraction of the Photometry Group 2 X-ray, 24 \micron\ and radio sources identified
as Variability Level 2 variables as a function of the  $[3.6]-[4.5]$ color.
\label{fig:col-col_Wave}}
\end{figure}

In Figure~\ref{fig:z_abs}, we present the dependence of variability of the AGES confirmed quasars on the 
redshift, absolute and observed magnitudes for both bands. The most variable objects seem to be those with redshifts $0.5<z<2$
and absolute magnitudes higher than $-25$. 

We have also matched our DIBS variability catalog to the XBo{\"o}tes X-ray catalog of the field (\citealt{2005ApJS..161....1M,2005ApJS..161....9K,2006ApJ...641..140B}),
MIPS detections at 24~\micron\ (from MAGES) and radio (FIRST; \citealt{1995ApJ...450..559B} and WSRT; \citealt{2002AJ....123.1784D}) catalogs.
We summarize the results in Table~\ref{tab:resultsWave} and in Figure~\ref{fig:col-col_Wave}.
AGNs are known to have X-ray emission, so it is not surprising to have a significant number of matches 
and that they mostly lie in the AGN color region (Figure~\ref{fig:col-col_Wave}; see also \citealt{2008ApJ...679.1040G}). 
Out of 39591 PG3 sources, we found 2307 matches to X-ray sources, of 
which 448 (19\%) were at least as variable as Variability Level 2. By increasing the variability threshold to VL4, 
we are left with 160 (6.9\%) X-ray-selected, mid-IR variable AGNs. Quasars are also known to have strong emission at mid- and far-IR. We matched
our variability catalog to MIPS sources detected at 24~\micron\ to find 26280 matches. Of them, 713 (2.7\%) were VL2 objects and 238 (0.9\%) were
VL4 sources. We also found nearly 1500 matches to radio catalogs. However, only 51 (3.4\%) objects were VL2 sources and there were only 17 (1.1\%) VL4 objects.

\cite{2014ApJ...790...54C} classified the Bo{\"o}tes sources based on up to 17-band SED fits using the \cite{2010ApJ...713..970A} 
templates and provided photometric redshifts for the sources lacking spectroscopy.
They used the $F$-test to compare fits with and without an AGN component,
where increasing values of $F$ make the presence of an AGN more probable.
Because non-AGNs greatly outnumber AGNs, there will be a high false positive rate for samples of objects
at $F$-test values that would strongly indicate the presence of an AGN if only considering a single object.
\cite{2014ApJ...790...54C} sources with $F>40$ almost all lie
in the AGN wedge, while those with $10<F<40$ track both the wedge and some of the more extended color region
where X-ray sources are found.
We have matched 3794 PG3 sources with $F>10$, of which 440 (11.6\%) turned out to exceed VL2 and 137 (3.6\%) to be VL4 sources.
These highly correlated variable sources also have very low estimated false positive rates of $\sim$1\%.
Of the 3794 PG3 sources, 2929 (77\%) lie inside the Stern wedge, 3282 (87\%) lie inside the modified wedge, and
only a single object (0\%) in the stellar box.

\begin{figure}
\centering
\includegraphics[width=8.2cm]{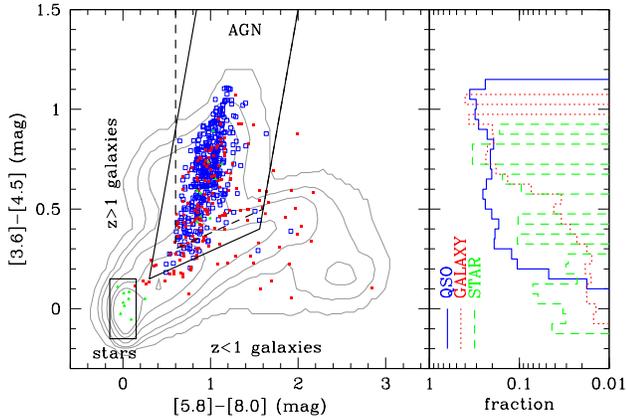}
\caption{
{\it Left panel:} Mid-IR color-color distribution of the Photometry Group 3, Variability Level 2 sources with spectroscopic classification
 from AGES.  Blue, red, and green symbols correspond to objects with quasar, galaxy, and stellar spectroscopic class, respectively.  
The contours and selection regions are the same as in Figure~\ref{fig:colcol_hist}.
 {\it Right panel:} The fraction of objects with stellar, galaxy or QSO template matches 
that are the Photometry Group 2, Variability Level 2 variables as a function of the  $[3.6]-[4.5]$ color. 
\label{fig:col-col_AGES}}
\end{figure}

The AGES survey provides spectroscopic redshifts and classifications for roughly 23,500 quasars, galaxies and stars (\citealt{2012ApJS..200....8K}). 
We matched the DIBS variability catalog to the AGES spectroscopic catalogs
and obtained 2179 matches for AGNs, 490 for stars, and 11199 for galaxies in Photometry Group 3 (Table~\ref{tab:resultsAGES}). At Variability Level 2,
we identified 402 (18.4\%) AGNs, 18 (3.7\%) stars, and 164 (1.5\%) galaxies (Figure~\ref{fig:col-col_AGES}). By increasing the variability significance threshold to VL4,
we were left with 145 (6.7\%) AGNs, 8 (1.6\%) stars, and 54 (0.5\%) galaxies.
 
\subsection{Comparison with SDWFS}

What are the implications of adding the fifth epoch (this study) to the SDWFS variability catalog?

Consider the Stern wedge sources first. The number of sources in the wedge is nearly identical in both surveys with 
6677 for SDWFS and 6692 for DIBS. There are 496 (7.4\%) SDWFS VL2 PG3 variable objects and 593 (8.9\%) DIBS sources.
DIBS increases the number of variable AGN candidates by nearly a hundred (20\%) and reduces the estimated false positive rate (from 6\% to 2\%).
Combining the two effects, the number of AGN with robustly measured variabilty is increased from roughly 470 to 581 or about 25\%.

Increasing the variability significance threshold to VL4 results in 204 (3.1\%) AGN in SDWFS 
and only 192 (2.9\%) in DIBS, and the false positive rate stays at 2\%.
On the other hand, if we keep the false positive rates from SDWFS (i.e., use a lower significance levels for DIBS), we
obtain 496 (SDWFS) and 936 (DIBS) PG3 sources for a 6\% false positive rate, 
or 204 (SDWFS) and 678 (DIBS) PG3 objects for a 2\% false positive rate.
The number of variable AGN detections is roughly doubled in DIBS as compared to SDWFS at fixed false positive rates.

We observe a similar situation for the modified AGN wedge. 
There are 584 (4.6\%) SDWFS VL2 sources and 652 (5.1\%) for DIBS (12\% more) out of 12741 and 12801 PG3 sources, respectively.
Along with the increased total number of these variable AGNs in DIBS, we observe a decreased false positive rate from 7\% to 3\%.
When increasing the variability significance threshold to VL4, we have 224 (1.8\%) and 209 (1.6\%) sources, respectively. 
For the fixed false positive rates from SDWFS we have 584/997 (SDWFS/DIBS) sources for a 7\% false positive rate or
224/719 for a 3\% false positive rate.

We also study the content of the stellar box, where there are 7567 (SDWFS) and 7477 (DIBS) objects.  While in SDWFS we 
have 86 (1.1\%) VL2 stars, there are 143 (1.9\%) in DIBS. Increasing the variability significance threshold to VL4 results
in a collection of 21 (0.3\%) stars in SDWFS and 68 (0.9\%) in DIBS.

\begin{deluxetable*}{cccccc}
\tablecaption{Mid-IR Variability of Photometrically Selected AGNs.\label{tab:resultsWave}}
\tablehead{
\makebox[1.4cm][c]{Variability} & 
\makebox[1.4cm][c]{Variability} &
\makebox[2.1cm][c]{X-ray} & 
\makebox[2.1cm][c]{FIRST} &
\makebox[2.1cm][c]{WSRT} &
\makebox[2.1cm][c]{\cite{2014ApJ...790...54C}}\\
Level & $\sigma_{12}$ & & & & 
}
\startdata
& & \multicolumn{4}{c}{Number (fraction/false positive rate) of variable objects: Photometry Group 1} \\
\cline{1-6} \\
Total & ---  &  3667              & 376               & 2300           & 9258  \\
2     & $>2$ &  469 (12.8\%/1\%)  & 11 (2.9\%/3\%)    & 42 (1.8\%/6\%) & 454 (4.9\%/3\%) \\
3     & $>3$ &  286 (7.8\%/1\%)   & 5 (1.3\%/2\%)     & 27 (1.2\%/6\%) & 254 (2.7\%/2\%) \\
4     & $>4$ &  163 (4.4\%/1\%)   & 4 (1.1\%/3\%)     & 14 (0.6\%/9\%) & 139 (1.5\%/2\%) \\
\\
& & \multicolumn{4}{c}{Number (fraction/false positive rate) of variable objects: Photometry Group 2} \\
\cline{1-6}\\
Total & ---  &   3258              & 342               & 2082           & 5550  \\
2     & $>2$ &   468 (14.4\%/1\%)  & 11 (3.2\%/3\%)    & 42 (2.0\%/6\%) & 451 (8.1\%/2\%) \\
3     & $>3$ &   286 (8.8\%/1\%)   & 5 (1.5\%/2\%)     & 27 (1.3\%/5\%) & 254 (4.6\%/1\%)  \\
4     & $>4$ &   163 (5.0\%/1\%)   & 4 (1.2\%/3\%)     & 14 (0.7\%/8\%) & 139 (2.5\%/1\%) \\
\\
& & \multicolumn{4}{c}{Number (fraction/false positive rate) of variable objects: Photometry Group 3} \\
\cline{1-6}\\
Total & ---  &   2307              & 206               & 1278           & 3794 \\
2     & $>2$ &   448 (19.4\%/1\%)  & 11 (5.3\%/2\%)    & 40 (3.1\%/1\%) & 440 (11.6\%/1\%) \\
3     & $>3$ &   277 (12.0\%/0\%)  & 5 (2.4\%/2\%)     & 25 (2.0\%/2\%) & 250 (6.6\%/1\%) \\
4     & $>4$ &   160 (6.9\%/0\%)   & 4 (1.9\%/3\%)     & 13 (1.0\%/4\%) & 137 (3.6\%/1\%)
\enddata
\tablecomments{The fraction of variable sources is a ratio of the number of variable objects above a given 
level of $\sigma_{12}$ and $r>0.8$ to the total number of objects in a respective class. The false positive rate is 
estimated from the number of sources above the same level of $\sigma_{12}$ but with $r<0.5$ (see \S\ref{sec:definitions} and Figure~\ref{fig:covariance}).}
\end{deluxetable*}

\begin{deluxetable*}{ccccc}
\tabletypesize{\scriptsize}
\tablecaption{Mid-IR Variability by AGES Spectroscopic Classification.\label{tab:resultsAGES}}
\tablewidth{0pt}
\tablehead{
\makebox[2.0cm][c]{Variability} & 
\makebox[2.5cm][c]{Variability} &
\makebox[2.5cm][c]{QSOs} & 
\makebox[2.5cm][c]{Stars} & 
\makebox[2.0cm][c]{Galaxies}\\
Level & $\sigma_{12}$ & & & 
}
\startdata
& & \multicolumn{3}{c}{Number (fraction/false positive rate) variable objects: Photometry Group 1} \\
\cline{1-5} \\
Total & ---  & 2911              & 1102                & 18039   \\
2     & $>2$ & 415 (14.3\%/1\%)  & 22 (2.0\%/16\%)     & 181 (1.0\%/16\%)  \\
3     & $>3$ & 245 (8.4\%/1\%)   & 13 (1.2\%/16\%)     & 110 (0.6\%/15\%)  \\
4     & $>4$ & 146 (5.0\%/1\%)   & 9 (0.8\%/10\%)      & 62 (0.3\%/21\%)   \\
\\
& & \multicolumn{3}{c}{Number (fraction/false positive rate) of variable objects: Photometry Group 2} \\
\cline{1-5} \\
Total & ---  & 2793              & 960                 & 17368  \\
2     & $>2$ & 415 (14.9\%/1\%)  & 22 (2.3\%/11\%)     & 178 (1.0\%/13\%)  \\
3     & $>3$ & 245 (8.8\%/1\%)   & 13 (1.4\%/10\%)     & 109 (0.6\%/14\%)  \\
4     & $>4$ & 146 (5.2\%/1\%)   & 9 (0.9\%/8\%)       & 62 (0.4\%/20\%)   \\
\\
& & \multicolumn{3}{c}{Number (fraction/false positive rate) of variable objects: Photometry Group 3} \\
\cline{1-5} \\
Total & ---  & 2179              & 490                 & 11199  \\
2     & $>2$ & 402 (18.4\%/1\%)  & 18 (3.7\%/5\%)      & 164 (1.5\%/7\%)  \\
3     & $>3$ & 240 (11.0\%/1\%)  & 11 (2.2\%/4\%)      & 100 (0.9\%/6\%)   \\
4     & $>4$ & 145 (6.7\%/1\%)   & 8 (1.6\%/3\%)       & 54 (0.5\%/9\%)  
\enddata
\tablecomments{The fraction of variable sources is a ratio of the number of variable objects above a given 
level of $\sigma_{12}$ and $r>0.8$ to the total number of objects in a respective class. The false positive rate is 
estimated from the number of sources above the same level of $\sigma_{12}$ but with $r<0.5$ (see \S\ref{sec:definitions} and Figure~\ref{fig:covariance}).}
\end{deluxetable*}


\section{Mid-IR Quasar Structure Functions}
\label{sec:sfunction}

Quasars are well known as aperiodically variable sources (\citealt{1997ARA&A..35..445U}). 
Such brightness changes are now routinely and successfully modeled using DRW model 
(e.g, \citealt{2009ApJ...698..895K,2010ApJ...708..927K,2010ApJ...721.1014M,2011AJ....141...93B,2011ApJ...728...26M,2011ApJ...735...80Z,2012ApJ...753..106M,2012ApJ...760...51R,2013ApJ...765..106Z}), 
however with some departures on very short time scales and low amplitudes that are presently only probed by {\it Kepler}
(\citealt{2011ApJ...743L..12M,2015MNRAS.451.4328K}).
Because our light curves are very sparsely sampled, we are restricted to comparing average structure functions 
(SF; e.g., \citealt{1998ApJ...504..671K,2004ApJ...601..692V,2005AJ....129..615D,2010ApJ...714.1194S}), which quantify the mean magnitude difference at 
the observed ($\tau=|t_j-t_i|$) or rest ($\tau=|t_j-t_i|/(1+z)$) frame
time difference (or ``lag'') between two epochs. We define it as in \cite{2010ApJ...716..530K},
\begin{eqnarray}
   S^2(\tau) & = & { 1 \over N_{qso}(\tau) } \sum_{i<j} \left( m(t_i) - m(t_j) \right)^2_{qso} - \nonumber  \\
             & - & { 1 \over N_{gal}(\tau) } \sum_{i<j} \left( m(t_i) - m(t_j) \right)^2_{gal},
\label{eqn:sf}
\end{eqnarray}
where $N_{qso}$ and $N_{gal}$ are the number of quasars and galaxies (providing the variance/noise estimate) used,
 and $m(t_i) - m(t_j)$ is the magnitude difference for a given time-lag $\tau$.
For each QSO, we match four galaxies having nearly identical brightness that serve as a noise estimate.
At the relevant magnitudes, galaxies are essentially unresolved in the DIBS images. 
We calculated 1000 bootstrap resamplings for both the quasar and galaxy lists, and we report here
the median values along with the 1$\sigma$ (68.3\%) uncertainties (Tables~\ref{tab:sf_fits_qso}--\ref{tab:sf_fits_chung}).
We do not expect a strong AGN variability signal at short time lags,
which makes these estimates very sensitive to any problems with the estimated noise. 
Having five epochs, we measure $n(n-1)/2=10$ different time-lags, blurred by the $(1+z)$ factor in the rest frame. 
We model the SF by a power-law of the form
\begin{equation}
     S(\tau) = S_0 \left( { \tau \over \tau_0 } \right)^\gamma,
\end{equation}
adopting $\tau_0=2$~(4) years for the rest-frame (observed-frame) estimates.

\begin{figure*}
\centering
\includegraphics[width=14cm]{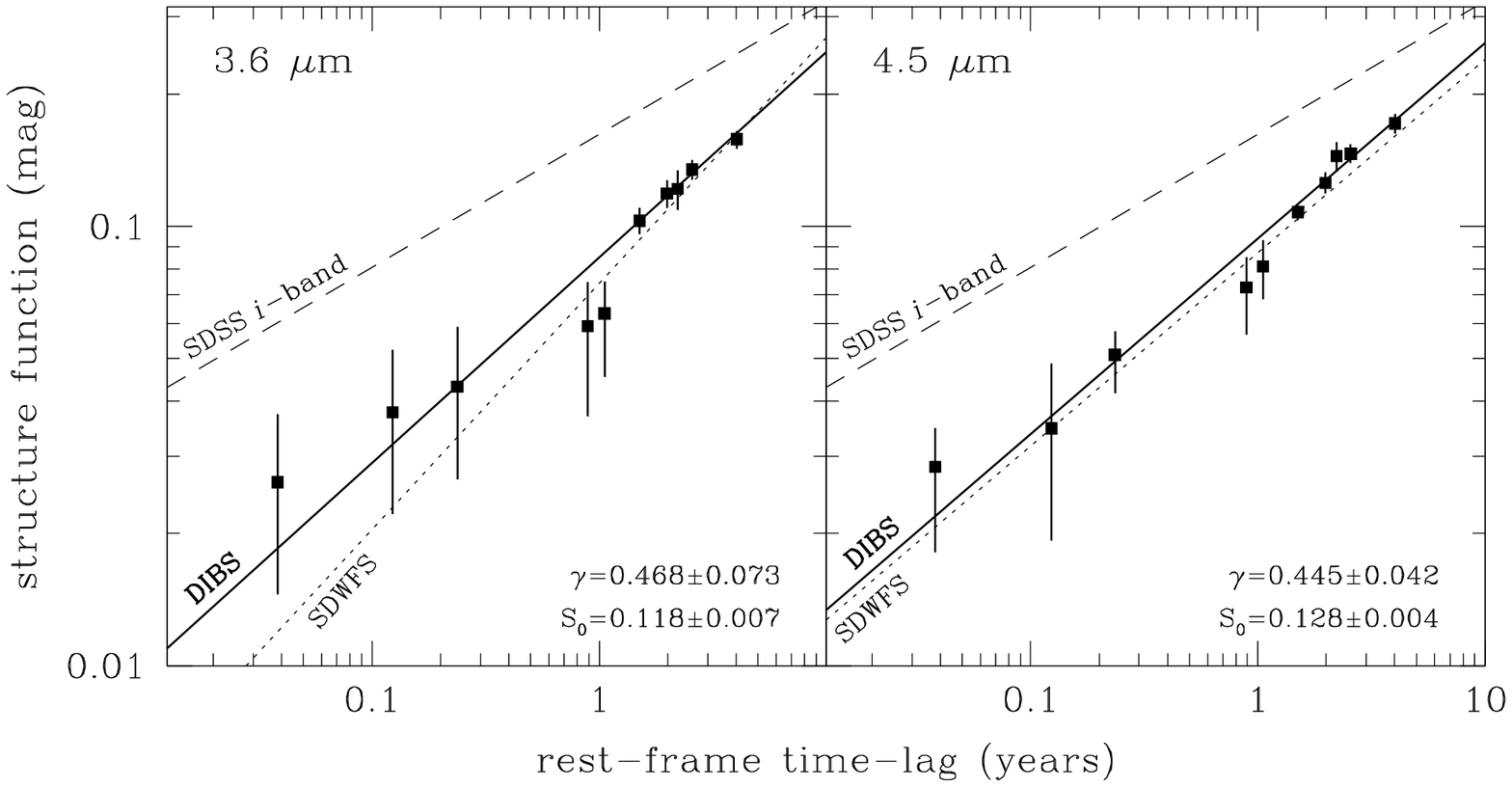}\\
\caption{Rest-frame mid-IR AGN structure functions for AGES quasars that are brighter than $[3.6]<18$ mag.
The solid lines represent the fit to the points. For comparison we also show
the mid-IR structure functions for SDWFS (dotted line; \citealt{2010ApJ...716..530K}) and
the optical $i$-band structure function for SDSS quasars (dashed line; \citealt{2004ApJ...601..692V}).
\label{fig:sf_bin}}
\end{figure*}

\begin{figure*}
\centering
\includegraphics[width=14cm]{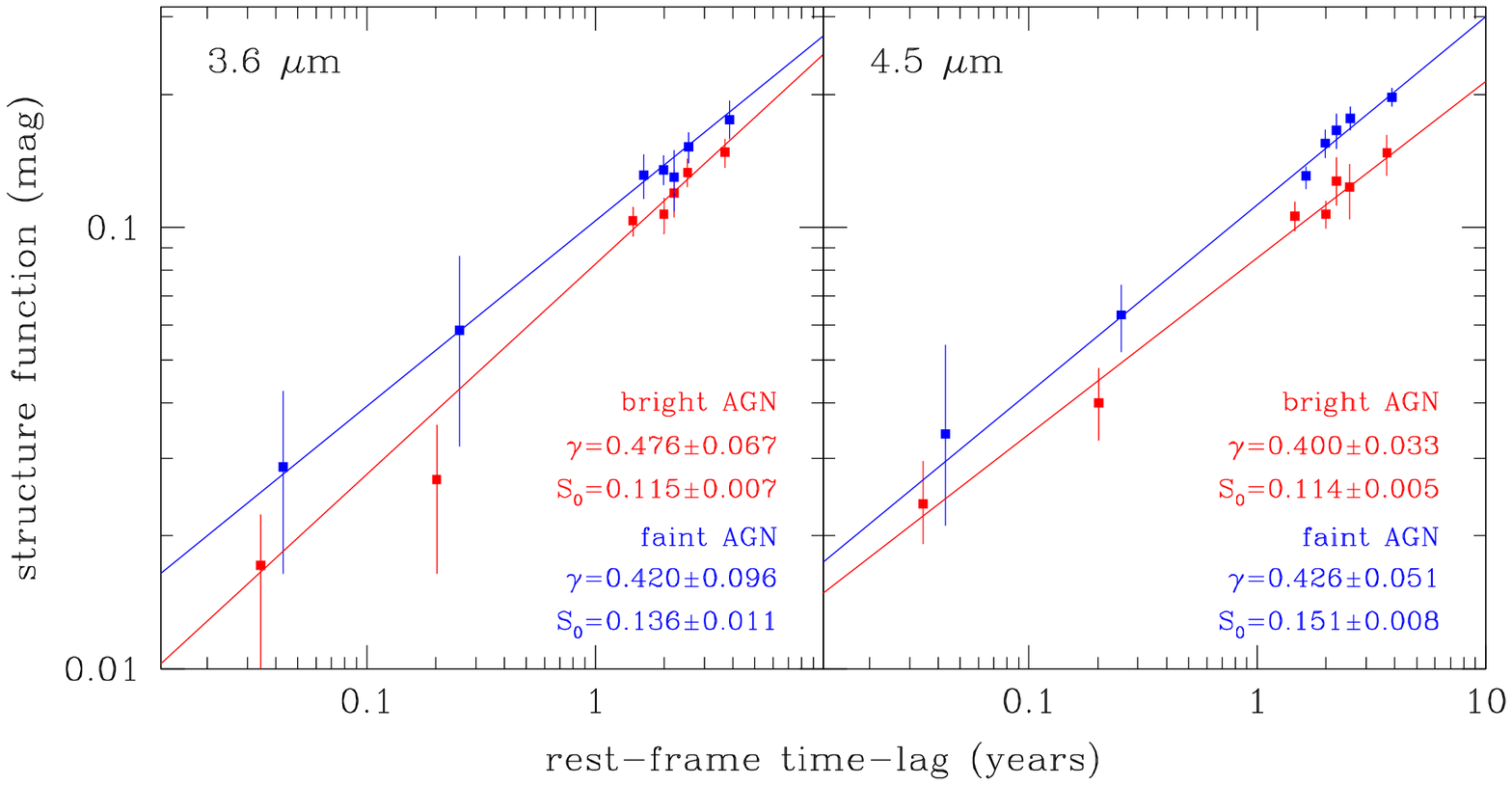}\\
\caption{Rest-frame mid-IR AGN structure functions for the $0.5<z<2$ AGES quasars brighter than $[3.6]<18$ mag.
We divide the sample of 1000 AGN in half after sorting them by their absolute magnitudes and calculate SF for the brighter half (red) and the fainter
half (blue). It is clear that fainter AGN  show higher amount of variability then the brighter ones.
\label{fig:sf_bightness}}
\end{figure*}

The AGES survey provides us with over 23500 sources in the field that have measured redshifts and/or type classification (quasar, galaxy, or star).
Because the number of galaxies in this magnitude limited survey falls off very quickly with increasing redshift, all sources  beyond the redshift $z>1$ are quasars
(see Figure~6 in \citealt{2012ApJS..200....8K}). We will frequently 
use this information, in the selection of ``true'' quasars and in the
calculation of the observed and rest-frame structure functions.

For the structure function, we must analyze both the detected and undetected variable AGNs in DIBS. We do this by computing 
the rest-frame structure function of all sources in a given sample with $[3.6]<$~18 mag. We consider several cases independent of their variability level.

\subsection{SF of the AGES quasars}

We calculated the rest-frame structure functions for spectroscopically confirmed quasars from AGES (Table~\ref{tab:sf_fits_AGESqso}).
The SF slopes are steeper than in the optical, with $\gamma=0.45$, and the amplitudes are $S_0=0.12$~mag at $\tau_0=2$~years (Figure~\ref{fig:sf_bin}), in 
agreement with our earlier findings in \cite{2010ApJ...716..530K}. The observed frame structure functions show slightly
shallower slopes with $\gamma\approx0.4$ and amplitude $S_0=0.11$~mag at $\tau_0=4$~years.

Let us now consider the dependence of the structure function on the redshift (or the emission wavelength) and the absolute luminosity of AGES quasars.
We select three redshift ranges of $0.7<z<1.2$, $1.2<z<1.5$, and $1.5<z<2.0$, each containing 248--336 AGNs.
The sample of quasars in each redshift bin was then divided into two equal halves after sorting them by absolute magnitude, hereafter called the faint and bright subsamples.
For the bright subsample, we observe constant SF slope across the three redshift bins (rest frame wavelengths $\lambda=1.2$--2.1$\mu$m) in both bands with mean $\gamma\approx0.44$.
We observe a mild decrease of the amplitude $S_0$ as a function of increasing redshift (correlation with wavelength). 

For the faint subsample, the SF slope is most likely constant at $\gamma\approx0.46$, but we observe a decline to $\gamma\approx0.37$ in both bands in the 
high-$z$ bin. From Figure~\ref{fig:z_abs}, we see
that with increasing redshift we are losing intrinsically faint AGNs (that seem to be more variable) due to the observational magnitude limit.
The observed decline of SF slope in the faint sample may simply reflect this fact, 
because we do not observe such a decline in the more complete bright subsamples.
For both the bright and faint subsamples, the amplitudes $S_0$ decline to higher redshifts and decreasing emission wavelengths.
Also fainter subsamples show larger amplitudes than for the bright samples. 

We repeated the SF calculation in equal magnitude--redshift
bins by dividing the AGES quasars into absolute magnitude bins (spaced every 1 mag)
and redshift bins (spaced every $\log{z}=0.2$), and considered only bins with a minimum of 10 AGNs.
The amplitude at a fixed time scale increases with increasing rest frame wavelength (or decreasing redshift). 
The slope is either constant with the rest frame wavelength (or redshift) or 
it is very weakly correlated with the rest frame wavelength (weakly anti-correlated with redshift).

To increase the statistical sample, we also define another
larger sample of 1000 AGNs with $0.5<z<2$. We again sort it by absolute brightness and split it into two brightness bins (Figure~\ref{fig:sf_bightness}). 
We do not find the slopes of the fainter AGNs to differ from the brighter ones (Figure~\ref{fig:sf_bightness}),
but the amplitudes are larger for fainter AGNs. This sample consists of 40\% highly correlated objects and 60\% with low or no correlation.
For the highly correlated objects, we again observe larger amplitudes for fainter AGNs, but also steeper SF slopes than for the bright AGNs. 

\subsection{SF of Mid-IR-selected Quasars}

We can also study the structure function of all AGN candidates lying inside the modified AGN color wedge. These sources must meet the PG3 requirement
but again with no restriction on their variability. There are 999
sources inside the wedge that are brighter than $[3.6]<18$~mag and at redshifts $z>1$. The latter requirement guarantees that we are dealing with true AGNs,
because galaxies are too faint to be observed at such redshifts in AGES.
The rest frame structure function slopes are $\gamma=0.43$--0.48 and amplitudes of $S_0=0.12$~mag (Table~\ref{tab:sf_fits_qso}), in agreement with the true AGES quasars.

We are also interested in the observed structure functions for $[3.6]<18$~mag PG3 sources inside the AGN color wedge with either $z>1$ (999 sources) or  
ignoring redshift information (11302 sources). 
The observed slopes for the first sample are $\gamma=0.42\pm0.08$ and $\gamma=0.39\pm0.04$ for the 3.6~\micron\ and 4.5~\micron\  bands. The 
amplitudes $S_0$ at the time scale $t_0=4$~years are $S_0=0.102\pm0.012$ mag and $S_0=0.107\pm0.008$ mag, respectively.
The second sample is characterized by SFs with $\gamma=0.37\pm0.12$ and $\gamma=0.34\pm0.08$ for the 3.6~\micron\ and 4.5~\micron\ bands, 
and the amplitudes of $S_0=0.064\pm0.012$ mag and $S_0=0.064\pm0.009$ mag, respectively.

\subsection{SF of X-ray-selected Quasars}

We analyzed the X-ray-matched sources with known redshifts from AGES that lie inside the modified AGN wedge.
We calculated the rest-frame structure functions  (Table~\ref{tab:sf_fits_Xray})
and obtained the following slopes $\gamma\approx0.46$ and amplitudes $S_0=0.13$~mag for AGNs at $z>1$, and obtained slopes 
$\gamma\approx0.44$ and amplitudes $S_0=0.14$~mag for all AGNs. These values again stay in general 
agreement with the measurements for the confirmed AGES quasars.

The observed structure functions for all $[3.6]<18$~mag PG3 X-ray sources inside the AGN color wedge, 
with or without known redshifts (1720 sources) have slopes
$\gamma=0.39\pm0.07$ and $\gamma=0.37\pm0.04$ for the 3.6~\micron\ and 4.5~\micron\ bands, and the 
amplitudes $S_0$ at the time scale $t_0=4$~years are $S_0=0.109\pm0.011$ mag and $S_0=0.122\pm0.008$ mag, respectively.

\subsection{SF of Radio-selected Quasars}

We also inspected the variability of 183 radio sources from inside the modified AGN wedge with known redshifts. We measure
the rest-frame slopes $\gamma=0.39\pm0.10$ for 3.6~\micron\ and $\gamma=0.36\pm0.08$ for 4.5~\micron\ bands. This selection returns
slightly shallower slopes than for the confirmed AGNs. The amplitudes at 2 years are $S_0=0.08$ and $S_0=0.09$~mag, respectively.
The observed structure functions for 555 sources from inside the modified AGN wedge with either known or unknown redshifts have slopes
$\gamma=0.32\pm0.08$ and $\gamma=0.36\pm0.09$ and amplitudes at the time scale $t_0=4$ years $S_0=0.060\pm0.008$ mag and $S_0=0.058\pm0.008$ mag,
 in the two bands, respectively.

\subsection{SF of Quasar-candidates from \cite{2014ApJ...790...54C}}

\cite{2014ApJ...790...54C} provide a catalog of AGN candidates based on up to 17-band SED fits using the \cite{2010ApJ...713..970A} templates. 
We have kept only the sources with $[3.6]<18$~mag, $F$-ratio
$F>10$ or $F>30$ for which we adopted the AGES spectroscopic redshifts and if unavailable then we used the estimated photometric redshifts. 
We find the rest-frame slopes $\gamma\approx0.37$ and
amplitudes $S_0\approx0.08$~mag for the $F>10$ sources, and
 $\gamma\approx0.40$ and
amplitudes $S_0\approx0.09$~mag for the $F>30$ sources.
These values are slightly lower than the ones for the confirmed AGNs.
This is most likely an indication of the level of contamination in the \cite{2014ApJ...790...54C} selection method.
In fact, we can use the amplitude of the structure function to
estimate the purity of the \cite{2014ApJ...790...54C} quasar candidate sample as a function of the $F$-test values.
We assume that the spectroscopically confirmed AGES AGN are representative
of AGN variability with $S_{\rm true}=0.12$ mag (0.13) mag at $t_0=2$ years and $\gamma=0.47$ (0.45) in the 3.6~\micron\ (4.5$\mu$m) band.

If we measure amplitude $S(F)$ (at fixed slope) for a sample where fraction
$f$ is composed of sources varying with an average amplitude of $S_{\rm true}$
and fraction $1-f$ are non-variable, then the fraction of real sources is
$f(F)= S(F)^2/S_{\rm true}^2$.   The scaling of $f(F)$ with $F$ is robust,
but the quantitative estimate depends strongly on using the correct
value of $S_{\rm true}$, particularly as $S(F)\rightarrow S_{\rm true}$. The
formal uncertainty is $\sigma_f = 2 f \sigma_{\rm true}/S_{\rm true}$ so
even our roughly $\sigma_{\rm true}/S_{\rm true} \simeq 5\%$ measurement
uncertainties represent a $10\%$ uncertainty in $f(F)$ as $f(F) \rightarrow 1$.

If we consider the \cite{2014ApJ...790...54C} sources in ``logarithmic'' bins
of $F=1-3$, $3-10$, $10-30$ and $>30$, we find 
$S(F)=0.025$, $0.029$, $0.070$ and $0.091$, respectively, at 3.6$\mu$m.  This then implies 
$f=0.04\pm0.01$, $0.06\pm 0.01$, $0.35\pm 0.05$ and
$0.59\pm 0.04$.  The results for 4.5~\micron\ are similar.  We clearly
see that the purity of the sample increases with $F$.  The samples
with $F<10$ are dominated by contaminants, as expected.  While
these values of $F$ would imply a significant AGN component when
considering a single object, they are not as significant when used
for a sample of objects where only $\sim 1\%$ of the sources are
AGN (see the discussion in \citealt{2014ApJ...790...54C}).   The low fraction
found for $F>30$ sources suggests that the uncertainties in $S_{\rm true}$
are dominated by systematic effects (e.g., are the AGES AGN representative?)
rather than the measurement errors.
The fraction $f(F)$ is, of course, a crude statistical estimate and likely
subject to a number of systematic effects.

\subsection{Optical vs. Mid-IR SFs}

From studies of optical quasar variability, we know that the structure function slopes are shallower at $\gamma\approx0.3$ (\citealt{2004ApJ...601..692V}) than
the mid-IR ones reported here.
In Figure~\ref{fig:sf_bin}, we compare the optical and mid-IR structure functions.
It is clear that at short time scales the variability amplitude in the mid-IR is significantly lower than that in optical, and the two match each other only at time-lags of several years. 
This is consistent with the wavelength dependent smoothing seen in NGC~2617 (\citealt{2014ApJ...788...48S}).


\begin{deluxetable}{lccr}
\tabletypesize{\scriptsize}
\tablecaption{Mid-IR Structure Functions for confirmed AGES Quasars ($[3.6]<18$~mag)\label{tab:sf_fits_AGESqso}}
\tablewidth{0pt}
\tablehead{ &  \multicolumn{2}{c}{Structure Function Parameters} & N$_{\rm obj}$\\
Band & \makebox[1.0cm][c]{$\gamma$} & \makebox[1.0cm][c]{$S_0$ (mag)} & }
\startdata
\\
\multicolumn{4}{c}{rest-frame ($\tau_0=2$~yrs)} \\
\cline{1-4}\\
3.6~\micron\ & $0.468 \pm 0.073$ &  $0.118 \pm 0.007$ & 1518 \\
4.5~\micron\ & $0.445 \pm 0.042$ &  $0.128 \pm 0.004$ & 1518 \\
3.6~\micron\ & fixed 0.5       &    $0.113 \pm 0.006$ & 1518 \\
4.5~\micron\ & fixed 0.5       &    $0.134 \pm 0.006$ & 1518 \\
\cline{1-4} \\
\\
\multicolumn{4}{c}{observed-frame ($\tau_0=4$~yrs)} \\
\cline{1-4} \\
3.6~\micron\ & $0.373 \pm 0.081$ &  $0.110 \pm 0.014$ & 1518  \\
4.5~\micron\ & $0.423 \pm 0.041$ &  $0.121 \pm 0.007$ & 1518
\enddata
\end{deluxetable}


\begin{deluxetable}{lccr}
\tablecaption{Mid-IR Structure Functions for AGN candidates inside the modified AGN wedge\\ (sources with $z>1$ and $[3.6]<18$~mag)\label{tab:sf_fits_qso}}
\tablewidth{0pt}
\tablehead{ & \multicolumn{2}{c}{Structure Function Parameters} & N$_{\rm obj}$ \\
Band &  \makebox[1.0cm][c]{$\gamma$} & \makebox[1.0cm][c]{$S_0$ (mag)} & }
\startdata
\\
\multicolumn{4}{c}{rest-frame ($\tau_0=2$~yrs)} \\
\cline{1-4}\\
3.6~\micron\ & $0.479 \pm 0.078$ &  $0.116 \pm 0.007$ & 999 \\
4.5~\micron\ & $0.425 \pm 0.043$ &  $0.122 \pm 0.004$ & 999 \\
3.6~\micron\ & fixed 0.5       &    $0.111 \pm 0.005$ & 999 \\
4.5~\micron\ & fixed 0.5       &    $0.122 \pm 0.003$ & 999 \\
\cline{1-4}
\\
\multicolumn{4}{c}{observed-frame ($\tau_0=4$~yrs)} \\
\cline{1-4}\\
3.6~\micron\ & $0.423 \pm 0.080$ &  $0.102 \pm 0.012$ & 999 \\
4.5~\micron\ & $0.392 \pm 0.042$ &  $0.107 \pm 0.008$ & 999 
\enddata
\end{deluxetable}


\begin{deluxetable}{lcccr}
\tabletypesize{\scriptsize}
\tablecaption{Mid-IR Structure Functions for X-ray-detected sources inside the modified AGN wedge ($[3.6]<18$~mag)\label{tab:sf_fits_Xray}}
\tablewidth{0pt}
\tablehead{ & & \multicolumn{2}{c}{Structure Function Parameters} & N\\
Band & $z>1$ &  \makebox[1.0cm][c]{$\gamma$} & \makebox[1.0cm][c]{$S_0$ (mag)} & }
\startdata
\\
\multicolumn{5}{c}{rest-frame ($\tau_0=2$~yrs)} \\
\cline{1-5} \\
3.6~\micron\ & Y & $0.463 \pm 0.093$ &  $0.125 \pm 0.009$ &  625 \\
4.5~\micron\ & Y & $0.458 \pm 0.042$ &  $0.126 \pm 0.005$ &  625 \\
3.6~\micron\ & Y & fixed 0.5       &  $0.128 \pm 0.008$ &  625 \\
4.5~\micron\ & Y & fixed 0.5       &  $0.130 \pm 0.010$ &  625 \\
3.6~\micron\ & N & $0.435 \pm 0.082$ &  $0.125 \pm 0.008$ & 1079 \\
4.5~\micron\ & N & $0.432 \pm 0.045$ &  $0.136 \pm 0.005$ & 1079 \\
3.6~\micron\ & N & fixed 0.5         &  $0.125 \pm 0.007$ & 1079 \\
4.5~\micron\ & N & fixed 0.5         &  $0.138 \pm 0.006$ & 1079 \\
\cline{1-5} \\
\\
\multicolumn{5}{c}{observed-frame ($\tau_0=4$~yrs)} \\
\cline{1-5} \\
3.6~\micron\ & Y & $0.412 \pm 0.058$ &  $0.112 \pm 0.012$ &  625 \\
4.5~\micron\ & Y & $0.430 \pm 0.047$ &  $0.111 \pm 0.009$ &  625 \\
3.6~\micron\ & N & $0.385 \pm 0.068$ &  $0.109 \pm 0.011$ & 1720 \\
4.5~\micron\ & N & $0.372 \pm 0.039$ &  $0.122 \pm 0.008$ & 1720
\enddata
\end{deluxetable}


\begin{deluxetable}{lccr}
\tabletypesize{\scriptsize}
\tablecaption{Mid-IR Structure Functions for $[3.6]<18$~mag AGN candidates \\from \cite{2014ApJ...790...54C} \label{tab:sf_fits_chung}}
\tablewidth{0pt}
\tablehead{ &  \multicolumn{2}{c}{Structure Function Parameters} & N\\
Band &  \makebox[1.0cm][c]{$\gamma$} & \makebox[1.0cm][c]{$S_0$ (mag)} & }
\startdata
\\
\multicolumn{4}{c}{$F>30$, rest-frame ($\tau_0=2$~yrs)} \\
\cline{1-4} \\
3.6~\micron\ & $0.431 \pm 0.081$ &  $0.087 \pm 0.007$ &  1518 \\
4.5~\micron\ & $0.384 \pm 0.077$ &  $0.106 \pm 0.008$ &  1518 \\
3.6~\micron\ & fixed 0.5 &  $0.084 \pm 0.005$ &  1518 \\
4.5~\micron\ & fixed 0.5 &  $0.100 \pm 0.005$ &  1518 \\
\cline{1-4} \\
\\
\multicolumn{4}{c}{$F>30$, observed-frame ($\tau_0=4$~yrs)} \\
\cline{1-4} \\
3.6~\micron\ & $0.395 \pm 0.050$ &  $0.086 \pm 0.008$ & 1518 \\
4.5~\micron\ & $0.381 \pm 0.048$ &  $0.096 \pm 0.008$ & 1518 \\
\\
\multicolumn{4}{c}{$F>10$, rest-frame ($\tau_0=2$~yrs)} \\
\cline{1-4} \\
3.6~\micron\ & $0.370 \pm 0.072$ &  $0.067 \pm 0.006$ &  4160 \\
4.5~\micron\ & $0.335 \pm 0.044$ &  $0.090 \pm 0.003$ &  4160 \\
3.6~\micron\ & fixed 0.5 &  $0.075 \pm 0.005$ &  4160 \\
4.5~\micron\ & fixed 0.5 &  $0.095 \pm 0.006$ &  4160 \\
\cline{1-4} \\
\\
\multicolumn{4}{c}{$F>10$, observed-frame ($\tau_0=4$~yrs)} \\
\cline{1-4} \\
3.6~\micron\ & $0.364 \pm 0.059$ &  $0.069 \pm 0.007$ & 4160 \\
4.5~\micron\ & $0.382 \pm 0.056$ &  $0.078 \pm 0.008$ & 4160
\enddata
\end{deluxetable}


\section{Summary}
\label{sec:summary}

In this paper, we have combined the images from SDWFS with a fifth epoch
from DIBS, extending the time baseline of the survey to 10 years.
We performed DIA photometry on the five epochs of DIBS data, providing the light curves
and variability measures for half a million mostly extragalactic sources.

We cross-correlated the DIBS variability data with the AGES redshift survey, X-ray, MIPS, radio data, and photometrically selected AGN candidates 
from the same area of the sky. In comparison to SDWFS, the variability yields are by 
15--25\% larger for X-ray and 24~\micron\ quasars, and nearly identical for radio sources exceeding the
variability significance $\sigma_{12}>2$. 
The yield of variable AGNs selected from the AGES survey seem to be also increased by $\sim$20\% as compared to SDWFS,
while the number of variable stars and galaxies stay the same. 
The mid-IR variability selection detects 18\% of the known AGNs from the AGES survey.
Keeping the false positive rates as in SDWFS (i.e., lowering the variability significance level used for the DIBS sample)
doubles the number of selected variable AGNs. The addition of the fifth epoch significantly lowered our estimated false positive rates.
The primary reason for the low overall yields is that quasar variability amplitudes are comparable (for $[3.6]<18$~mag) to the photometric quality of the DIBS data.

The main motivation of this paper was to study the variability parameters (the SF slope and amplitude) as a function
of physical quantities (luminosity or wavelength) or selection method (true AGNs, or mid-IR, X-ray, or radio selected, or from SED fitting).
The variability analysis of $\sim$1500 spectroscopically confirmed sources from the AGES survey returned 
the rest-frame structure function slopes in both bands of $\gamma\approx0.45$ and amplitudes $S_0=0.11$~mag (at $\tau_0=2$~years).
Except for the small sample of radio-selected AGNs the mid-IR structure functions of AGNs
are significantly steeper than in optical ($\gamma\approx0.3$).
The optical and mid-IR variability amplitudes match each other for time-lags of several years, but for the shorter lags the mid-IR variability amplitudes
are significantly lower than the ones in optical. 
The lack of short time scale variability can be interpreted as the accretion disk or dust torus smoothing out short-time scale variations. 
This is easily understood, as the accretion disk radius is $R_\lambda\approx 4\times(\lambda_{\rm rest}/{\rm\mu m})^{4/3}$~light days for a typical black hole mass of $10^9$ solar masses,
radiative efficiency of 0.1, and Eddington ratio of $1/3$, hence the optical variations originate in a region of 1 light-day in size and the mid-IR ones from a region 25 times bigger.

We also explored the dependence of the structure function slope and amplitude as a function of the luminosity and rest-frame wavelength.
We find significant anti-correlation (lack of correlation) of the AGN amplitude (slope) with luminosity. In accordance with optical studies, the brighter the AGN
the lower its variability amplitude. We also find no clear evidence for the correlation of the variability slope with the rest-frame wavelength, but the 
range of rest-frame wavelengths explored is rather narrow ($\lambda=1.2$--2.1$\mu$m).
On the other hand, there is a correlation of the SF amplitude with the rest-frame wavelength.


\acknowledgments
We thank Prof. Igor Soszy{\'n}ski for discussions on variable stars
and the anonymous referee for helpful suggestions that improved the manuscript.
This work is based on observations made with the {\it Spitzer Space Telescope}, 
which is operated by the Jet Propulsion Laboratory,
California Institute of the Technology under contract
with the National Aeronautics and Space Administration (NASA).
SK acknowledges the financial support of the Polish National Science
Center through the grant number 2014/15/B/ST9/00093.
RJA was supported by Gemini-CONICYT grant number 32120009 and FONDECYT grant number 1151408.

Facilities: \facility{{\it Spitzer Space Telescope} (IRAC)}


\end{document}